\documentclass[jcp,twocolumn,superscriptaddress,10pt]{revtex4}
\usepackage{graphicx}
\usepackage{subfigure}
\usepackage{amsmath}
\usepackage{amsfonts}
\usepackage{amssymb}
\usepackage{amsthm}
\usepackage{times}
\usepackage[colorlinks,citecolor=blue,linkcolor=blue]{hyperref}
\usepackage{color}
\usepackage{setspace}

\begin{document}
\title{Heat transfer statistics in mixed quantum-classical systems}

\newcommand{\UA}{\affiliation{Department of Chemistry, University of Alberta, Edmonton, Alberta, T6G 2G2, Canada }}

\newcommand{\UT}{\affiliation{Department of Chemistry and Centre for Quantum Information and Quantum Control, University of Toronto, 80 Saint George St., Toronto, Ontario, M5S 3H6, Canada }}

\newcommand{\SM}{\affiliation{Singapore-MIT Alliance for Research and Technology (SMART) center, 1 CREATE Way, Singapore 138602, Singapore}}

\author{Junjie Liu}
\UA
\author{Chang-Yu Hsieh}
\SM
\author{Dvira Segal}
\UT
\author{Gabriel Hanna}
\email{gabriel.hanna@ualberta.ca}
\UA

\begin{abstract}
The modelling of quantum heat transfer processes at the nanoscale is crucial for the development of energy harvesting and molecular electronics devices. Herein, we adopt a mixed quantum-classical description of a device, in which the open subsystem of interest is treated quantum mechanically and the surrounding heat baths are treated in a classical-like fashion.  By introducing such a mixed quantum-classical description of the composite system, one is able to study the heat transfer between the subsystem and bath from a closed system point of view, thereby avoiding simplifying assumptions related to the bath time scale and subsystem-bath coupling strength. In particular, we adopt the full counting statistics approach to derive a general expression for the moment generating function of heat in systems whose dynamics are described by the quantum-classical Liouville equation (QCLE). From this expression, one can deduce expressions for the dynamics of the average heat and heat current, which may be evaluated using numerical simulations.  Due to the approximate nature of the QCLE, we also find that the steady state fluctuation symmetry holds up to order $\hbar$ for systems whose subsystem-bath couplings and baths go beyond bilinear and harmonic, respectively.  To demonstrate the approach, we consider the nonequilibrium spin boson model and simulate its time-dependent average heat and heat current under various conditions. 
\end{abstract}

\maketitle

\section{Introduction}
Owing to the rapid development of nanotechnologies in recent decades, heat transfer at the nanoscale has attracted significant attention. Numerous studies have been dedicated to gaining a deep understanding and precise control of the heat transfer, which has impacts at both the fundamental and practical levels. So far, heat transfer has been studied in small and well-characterized quantum systems. On the experimental side, systems such as molecular junctions can be fabricated in the laboratory \cite{Wang.07.S,Schwab.00.N,Carter.09.ACR,Mark.12.NM,Meier.14.PRL,Cui.17.S}, while on the theoretical side, simplified models can be put forward and studied with a host of fully quantum methods \cite{Segal.05.PRL,Velizhanin.08.CPL,Ren.10.PRL,Nicolin.11.JCP,Nicolin.11.PRB,Ruokola.11.PRB,Segal.13.PRB,Saito.13.PRL,Yang.14.EL,Wang.15.SR,Carrega.16.PRL,Liu.17.PRE,Wang.17.PRA}.

When the heat transfer occurs in a complex, many-body system such as a molecular aggregate \cite{May.11.NULL} or a self-assembled monolayer junction \cite{Majumdar.15.NL}, which may not be well described in terms of a simplified model containing a small number of degrees of freedom (DOF), a fully quantum approach to modelling the heat transfer dynamics will be computationally intractable. In this case, an approximate treatment of the dynamics is required to gain insight into the system under study.  Mixed quantum-classical dynamics methods, which treat a set of light particles of interest (i.e., subsystem) quantum mechanically and the remaining particles in the system (i.e., bath or environment) in a classical-like fashion, provide tremendous computational advantages over fully quantum methods \cite{Tully.90.JCP,Prezhdo.97.PRA,Martens.97.JCP,Tully.98.FD,Kapral.99.JCP,Wan.00.JCP,Horenko.02.JCP,Kelly.13.JCP,Bai.14.JPCA,Kim.14.JCP,Wang.15.JPCL,Martens.16.JPCL,Wang.16.JPCL,Agostini.16.JCTC,Subotnik.16.ARPC}. 

In this work, we adopt a mixed quantum-classical approach to modelling heat transfer dynamics that is based on the quantum-classical Liouville equation (QCLE) \cite{Aleksandrov.81.ZNA,Gerasimenko.82.TMP,Zhang.88.JPP,Kapral.99.JCP}, which stems from a linearization of the quantum Liouville equation expressed in the partial Wigner representation \cite{Wigner.32.PR}, viz., a description of the subsystem and bath DOF in terms of operators and phase space variables, respectively.  The QCLE is chosen as the starting point for our work because (i) several of the popular mixed quantum-classical methods may be derived from this equation \cite{Kapral.15.JP,Kapral.16.CP}, and (ii) it yields the exact quantum dynamics for quantum subsystems that are bilinearly coupled to harmonic environments \cite{Kernan.02.JCP}, which are frequently used as models for studying energy transfer at the nanoscale.  In particular, we combine the QCLE and full counting statistics (FCS) \cite{Levitov.93.JETP,Levitov.96.JMP,Belzig.01.PRL,Klich.03.NULL,Bagrets.03.PRB,Pilgram.03.PRL,Saito.08.PRB,Gutman.10.PRL} approaches to derive a general expression for the moment generating function (MGF) of heat, which may then be used to compute the time-dependent average heat and its fluctuations in a system. As the QCLE treats the dynamics of the heat baths explicitly, one can start from the exact definition of the MGF in FCS and does not need to impose any constraints on the bath timescale and subsystem-bath coupling strength, in contrast to the conventional Redfield master equation \cite{Segal.05.PRL} and nonequilibrium Green's function method \cite{Liu.17.PRE}.  Thus, one can apply this combined approach to a wide range of parameter regimes. 

Because heat fluctuates at the nanoscale, its average is insufficient to fully characterize a heat transfer process. For a fully quantum system at steady state, heat fluctuations are governed by the steady state fluctuation symmetry (SSFS) of the MGF \cite{Esposito.09.RMP,Campisi.11.RMP,Nicolin.11.JCP}. However, when the dynamics of a fully quantum system is approximated, the behavior of the heat fluctuations may be altered and, as a result, the SSFS may not be satisfied.  A direct consequence of this is the breakdown of the fluctuation-dissipation theorem in the linear response regime. Thus, it is of vital importance to assess to what extent the SSFS holds in systems whose dynamics are described by the QCLE. In the case of systems for which QCLE dynamics is exact (e.g., subsystems that are bilinearly coupled to harmonic environments), one expects the SSFS to be strictly preserved, while in the case of systems for which QCLE dynamics is approximate, one expects to reach an approximate nonequilibrium steady state. Nevertheless, in the limit of high temperature and a very small mass ratio between the subsystem and bath particles, the approximations introduced by the QCLE dynamics are expected to be minor. 

To illustrate the utility of our approach, we consider the nonequilibrium spin-boson (NESB) model, a prototypical model in the study of quantum energy transfer over the last decade \cite{Boudjada.14.JPCA}. In particular, we compute the time-dependent average heat and heat current using a recently proposed method for solving the QCLE \cite{Liu.18.NULL}.  This method deterministically propagates the dynamics of the system by numerically solving a set of coupled first-order differential equations for the subsystem and bath coordinates.  Given its demonstrated accuracy and efficiency in several prototype systems, we believe that a QCLE-based approach to heat transfer statistics will provide a viable way of studying more realistic models of many-body systems.

The paper is organized as follows. We describe the model and MGF of heat in section \ref{sec:2}. In section \ref{sec:3}, we derive a general expression for the MGF of heat in the quantum-classical limit. In section \ref{sec:4}, we address the question of the extent to which the SSFS holds in systems whose dynamics are described by the QCLE.  In section \ref{sec:5}, we apply our formalism to the NESB model and present and discuss our numerical results for the time-dependent heat and heat current. We summarize our findings in section \ref{sec:6}.

\section{General background}\label{sec:2}
\subsection{Model}
We consider a composite quantum system in which a subsystem is in contact with $K$ ($K\geq2$) bosonic heat baths at different temperatures and whose Hamiltonian is given by 
\begin{equation}\label{eq:hh}
\hat{H}~=~\hat{H}_S(\boldsymbol{\hat{x}})+\sum_{v=1}^K\hat{H}_{B}^v(\boldsymbol{\hat{X}}_v)+\hat{H}_I(\boldsymbol{\hat{x}},\boldsymbol{\hat{X}}),
\end{equation}
where $\hat{H}_S$ is the subsystem Hamiltonian; $\hat{H}_B^v=\sum_{j=1}^{N_v}[\hat{P}_{j,v}^2/2+\omega_{j,v}^2\hat{R}_{j,v}^2/2]$ is the Hamiltonian of the $v$th heat bath at inverse temperature $\beta_v$ with $\hat{P}_{j,v}$, $\hat{R}_{j,v}$, and $\omega_{j,v}$ the mass-weighted momentum, position, and frequency of the $j$th oscillator, respectively; and $\hat{H}_I$ is the subsystem-bath interaction Hamiltonian with $\boldsymbol{\hat{X}}=(\boldsymbol{\hat{X}}_1,\boldsymbol{\hat{X}}_2,\ldots,\boldsymbol{\hat{X}}_K)$. In the above equation, $\boldsymbol{\hat{x}}=(\boldsymbol{\hat{r}},\boldsymbol{\hat{p}})$ and $\boldsymbol{\hat{X}}_v=(\boldsymbol{\hat{R}}_v,\boldsymbol{\hat{P}}_v)$ with $\boldsymbol{\hat{R}}_v=(\hat{R}_{1,v},\hat{R}_{2,v},\ldots,\hat{R}_{N_v,v})$ and $\boldsymbol{\hat{P}}_v=(\hat{P}_{1,v},\hat{P}_{2,v},\ldots,\hat{P}_{N_v,v})$, where $N_v$ is the number of harmonic oscillators in the $v$th heat bath. In what follows, we assume factorized initial density operators $\hat{\rho}_0=\hat{\rho}_S(0)\otimes\hat{\rho}_B(0)$, where $\hat{\rho}_S(0)$ is the initial subsystem density operator and $\hat{\rho}_B(0)=\hat{\rho}_B^1(0)\otimes\cdots\otimes\hat{\rho}_B^K(0)$ is the initial bath density operator with each $\hat{\rho}_B^v(0)\propto e^{-\beta_v\hat{H}_B^v}$ assuming a canonical form. 

To quantify the heat transfer between the subsystem and its heat baths, we define the average of the heat transferred from the $v$th heat bath to the subsystem as the average change in the bath energy during a time interval $[0,t]$ \cite{Bijay.12.PRE} 
\begin{equation}\label{eq:q_definition}
\langle Q_v(t)\rangle~=~\langle\hat{H}_B^v(0)-\hat{H}_B^v(t)\rangle,
\end{equation}
where the time dependence should be understood in the Heisenberg picture.  It follows that the average heat current from the $v$th heat bath to the subsystem may be obtained by taking the time derivative of the above equation, i.e., 
\begin{equation}\label{eq:j_definition}
\langle J_v(t)\rangle~=~-\frac{d}{dt}\langle \hat{H}_B^v(t)\rangle.
\end{equation}

In anticipation for a mixed quantum-classical description of the system's dynamics, we express Eqs.~(\ref{eq:q_definition}) and (\ref{eq:j_definition}) in the partial Wigner representation by taking the Wigner transform \cite{Wigner.32.PR} of these equations over the bath degrees of freedom (DOFs).   For a general operator $\hat{A}(t)$, its expectation value in this representation is given by $\langle \hat{A}(t) \rangle = \sum_{\alpha\alpha^{\prime}}\int d\boldsymbol{X}A_W^{\alpha\alpha^{\prime}}(\boldsymbol{X},t)\rho_{W}^{\alpha^{\prime}\alpha}(\boldsymbol{X},0)$, where $\{|\alpha\rangle\}$ denotes a complete set of basis states that span the Hilbert space of the subsystem, $(\cdot)_W^{\alpha\alpha^{\prime}}\equiv\langle \alpha |(\cdot)_W|\alpha^{\prime}\rangle$, and $\hat{\rho}_W(0)$ is the partial Wigner transform of $\hat{\rho}_0$ \cite{Sergi.03.TCA}. Using this result, one can directly write Eqs.~(\ref{eq:q_definition}) and (\ref{eq:j_definition}) in the partial Wigner representation as 
\begin{eqnarray}
\langle Q_v(t)\rangle &=& \sum_{\alpha\alpha^{\prime}}\int d\boldsymbol{X}\left[H_{B,W}^v(\boldsymbol{X})\delta_{\alpha\alpha^{\prime}}-(H_{B,W}^v)^{\alpha\alpha^{\prime}}(\boldsymbol{X},t)\right]\nonumber\\
 &&\times \rho_W^{\alpha^{\prime}\alpha}(\boldsymbol{X},0),\label{eq:heat}\\
 \langle J_v(t)\rangle &=&-\sum_{\alpha\alpha^{\prime}}\int d\boldsymbol{X}\left(\frac{d}{d t}H_{B,W}^v(\boldsymbol{X},t) \right)^{\alpha\alpha^{\prime}}\nonumber \\
 &&\times \rho_W^{\alpha^{\prime}\alpha}(\boldsymbol{X},0).\label{eq:hc}
\end{eqnarray}
In Eq.~(\ref{eq:heat}), the delta function results from the fact that the subsystem and bath are uncorrelated initially. However, at finite times, one must consider the matrix elements of the bath Hamiltonian in the subsystem basis because they depend on the subsystem operators due to the subsystem-bath interaction.

\subsection{Moment generating function of heat}
To fully characterize a heat transfer process at the nanoscale, not only is information about the average heat and heat current important, but one should also consider the higher order heat fluctuations. The FCS approach \cite{Levitov.93.JETP,Levitov.96.JMP,Belzig.01.PRL,Klich.03.NULL,Bagrets.03.PRB,Pilgram.03.PRL,Saito.08.PRB,Gutman.10.PRL} provides a general route for obtaining such statistics of heat in open quantum systems. Recalling that the heat transferred from the $v$th bath to the subsystem in a time interval $t$ is given by $\langle Q_v(t)\rangle=\langle \hat{H}^v_B(0)-\hat{H}^v_B(t)\rangle$, one may evaluate $\langle Q_v(t)\rangle$ using a two-time measurement \cite{Campisi.11.RMP} in which the instantaneous eigenvalues (eigenvectors) of $\hat{H}_B^v$ at time $t$ are $a_t$ ($|a_t\rangle$).  This two-time measurement can be described in terms of the joint probability of measuring $a_0$ at time zero and $a_t$ at time $t$, i.e., 
\begin{equation}\label{eq:p0t}
P(a_t,a_0)~=~\mathrm{Tr}\{\hat{\mathcal{P}}_{a_t}\hat{U}(t,0)\hat{\mathcal{P}}_{a_0}\hat{\rho}_0\hat{\mathcal{P}}_{a_0}U^{\dagger}(t,0)\hat{\mathcal{P}}_{a_t}\},
\end{equation}
where $\hat{\mathcal{P}}_{a_t}=|a_t\rangle\langle a_t|$, $\hat{U}(t,0)$ is the time evolution operator governed by the total Hamiltonian $\hat{H}$, and $\hat{\rho}_0$ is the initial total density operator. 
Since it has been previously shown that only the part of $\hat{\rho}_0$ that commutes with $\hat{H}_B^v$ determines the moment generating function \cite{Esposito.09.RMP}, for convenience, we choose $\hat{\rho}_0$ such that $[\hat{\rho}_0, \hat{H}_B^v]=0$ at time zero (which is the case for factorized initial states). 
Furthermore, since $[\hat{\mathcal{P}}_{a_0}, \hat{\rho}_0]=0$ (as a result of $[\hat{\rho}_0, \hat{H}_B^v]=0$), Eq.~(\ref{eq:p0t}) becomes
\begin{equation}\label{eq:p0t2}
P(a_t,a_0)~=~\mathrm{Tr}\{\hat{\rho}_0\hat{\mathcal{P}}_{a_0}\hat{U}^{\dagger}(t,0)\hat{\mathcal{P}}_{a_t}\hat{U}(t,0)\}.
\end{equation}

The probability distribution for the difference $\Delta a$ (i.e., the amount of heat transferred from the measured bath to the subsystem) between the output of the two aforementioned measurements is given by
\begin{equation} \label{eq:pt}
P_t(\Delta a)~=~\sum_{a_t,a_0}\delta(\Delta a-(a_0-a_t))P(a_t,a_0).
\end{equation}
The corresponding MGF may be defined as
\begin{equation}\label{eq:ztd}
Z(\chi_v,t)~\equiv~\int\,d\Delta a e^{i\chi_v\Delta a}P_t(\Delta a),
\end{equation}
where $\chi_v$ is the counting field associated with the measurement on the $v$th bath. Upon substituting Eq.~(\ref{eq:pt}) into Eq.~(\ref{eq:ztd}), the MGF becomes  
\begin{equation}\label{eq:ztd1}
Z(\chi_v,t)~=~\sum_{a_t,a_0} e^{-i\chi_v(a_t-a_0)}P(a_t,a_0).
\end{equation}
Noting that $f(\hat{B})=\sum_b\hat{\mathcal{P}}_bf(b)$, where $f$ is an arbitrary function of an arbitrary operator $\hat{B}$ with $\hat{B}|b\rangle=b|b\rangle$ and $\hat{\mathcal{P}}_b=|b\rangle\langle b|$, and substituting Eq.~(\ref{eq:p0t2}) into Eq.~(\ref{eq:ztd1}), the MGF simplifies to
\begin{equation}\label{eq:ztd4}
Z(\chi_v,t)~=~\mathrm{Tr}[e^{i\chi_v \hat{H}_B^v}e^{-i\chi_v \hat{H}_B^v(t)}\hat{\rho}_0],
\end{equation}
where $\hat{H}_B^v(t)=\hat{U}^{\dagger}(t,0)\hat{H}_B^v \hat{U}(t,0)$. Generalizing this expression to the multiple bath case leads to
\begin{equation}\label{eq:zt_total}
Z(\{\chi_v\},t)~=~\mathrm{Tr}\left[e^{i\sum\limits_{v}\chi_v \hat{H}_B^v}e^{-i\sum\limits_{v}\chi_v \hat{H}_B^v(t)}\hat{\rho}_0\right],
\end{equation}
where $\chi_v$ is the counting field for the $v$th heat bath, $\{\chi_v\}\equiv\{\chi_1,\chi_2,\ldots,\chi_K\}$, and the trace is performed over all DOFs.  It should be noted that, to arrive at this expression, one requires that $[\hat{\rho}_0, \hat{H}_B^v]=0$ \cite{Esposito.09.RMP}, which is the case for the factorized initial state $\hat{\rho}_0$.

By differentiating the MGF with respect to the counting field and evaluating the result at $\chi_v=0$, one obtains the $n$th moment of heat for the $v$th bath, i.e.,  
\begin{equation}\label{eq:mq}
\langle Q_v^n(t)\rangle=\left.\frac{\partial^n}{\partial(i\chi_v)^n}Z(\{\chi_v\},t)\right|_{\{\chi_v\}=0}.
\end{equation}
As $\langle Q_v(t)\rangle$ corresponds to the transferred energy from the $v$th bath to the subsystem during the time interval $[0,t]$, the time derivative of the first moment will give rise to the time-dependent energy current. Higher moments contain information about higher order correlations of the transferred energy. 

\section{Quantum-classical limit of MGF}\label{sec:3}
\subsection{Derivation of the MGF}
To obtain the quantum-classical limit of the MGF, we start by introducing a coordinate representation $\{\boldsymbol{\mathcal{Q}}\}=\{\boldsymbol{r},\boldsymbol{R}\}$ (calligraphic symbols are used to denote variables for the entire system) into Eq.~(\ref{eq:zt_total})
\begin{eqnarray}
Z(\{\chi_v\},t) &=& \int d\boldsymbol{\mathcal{Q}}_1d\boldsymbol{\mathcal{Q}}_2d\boldsymbol{\mathcal{Q}}_3d\boldsymbol{\mathcal{Q}}_4\langle \boldsymbol{\mathcal{Q}}_1|e^{i\sum_v\chi_v \hat{H}_B^v}|\boldsymbol{\mathcal{Q}}_2\rangle\nonumber\\
&&\times \langle\boldsymbol{\mathcal{Q}}_2|\hat{U}^{\dagger}(t,0)|\boldsymbol{\mathcal{Q}}_3\rangle\langle \boldsymbol{\mathcal{Q}}_3|e^{-i\sum_v\chi_v \hat{H}_B^v}|\boldsymbol{\mathcal{Q}}_4\rangle \nonumber\\
&&\times \langle\boldsymbol{\mathcal{Q}}_4|\hat{U}(t,0)\hat{\rho}_0|\boldsymbol{\mathcal{Q}}_1\rangle,
\end{eqnarray}
where $\hat{U}(t,0)$ is the time evolution operator governed by the total Hamiltonian $\hat{H}$.
We next make a change of variables, $\boldsymbol{\mathcal{Q}}_1=\boldsymbol{\mathcal{R}}_1-\boldsymbol{\mathcal{Z}}_1/2$, $\boldsymbol{\mathcal{Q}}_2=\boldsymbol{\mathcal{R}}_1+\boldsymbol{\mathcal{Z}}_1/2$, $\boldsymbol{\mathcal{Q}}_3=\boldsymbol{\mathcal{R}}_2-\boldsymbol{\mathcal{Z}}_2/2$, and $\boldsymbol{\mathcal{Q}}_4=\boldsymbol{\mathcal{R}}_2+\boldsymbol{\mathcal{Z}}_2/2$, and rewrite the above equation as
\begin{eqnarray}\label{eq:zt_q}
Z(\{\chi_v\},t) 
&=& \int d\boldsymbol{\mathcal{R}}_1d\boldsymbol{\mathcal{R}}_2d\boldsymbol{\mathcal{P}}_1d\boldsymbol{\mathcal{P}}_2D(\boldsymbol{\mathcal{R}}_1,\boldsymbol{\mathcal{P}}_1,\boldsymbol{\mathcal{R}}_2,\boldsymbol{\mathcal{P}}_2,t)\nonumber\\
&&\times \left(e^{i\sum_v\chi_v H_B^v}\right)_W(\boldsymbol{\mathcal{R}}_1,\boldsymbol{\mathcal{P}}_1)\nonumber\\
&&\times \left(e^{-i\sum_v\chi_v H_B^v}\right)_W(\boldsymbol{\mathcal{R}}_2,\boldsymbol{\mathcal{P}}_2),
\end{eqnarray}
where we have used the notation $\boldsymbol{\mathcal{R}}=(\boldsymbol{r},\boldsymbol{R})$ and $\boldsymbol{\mathcal{P}}=(\boldsymbol{p},\boldsymbol{P})$ (the lowercase and uppercase symbols refer to the subsystem and bath variables, respectively). To arrive at the above equation, we used the fact that the matrix element of an arbitrary operator $\hat{O}$ may be expressed in terms of its Wigner transform $O_W$ as follows
\begin{eqnarray}
&&\left\langle \boldsymbol{\mathcal{R}}-\frac{\boldsymbol{\mathcal{Z}}}{2}\left|\hat{O}\right|\boldsymbol{\mathcal{R}}+\frac{\boldsymbol{\mathcal{Z}}}{2}\right\rangle\nonumber\\
&&= \frac{1}{(2\pi\hbar)^{\mu}}\int d\boldsymbol{\mathcal{P}}e^{-(i/\hbar)\boldsymbol{\mathcal{P}}\cdot\boldsymbol{\mathcal{Z}}}O_W(\boldsymbol{\mathcal{R}},\boldsymbol{\mathcal{P}}),
\end{eqnarray}  
where $\mu=\mu_S+\mu_B$ is the coordinate-space dimension of the total system and $\boldsymbol{\mathcal{Z}}=(\boldsymbol{z},\boldsymbol{Z})$.  Finally, the time-dependent weight function $D$ has the following form
 \begin{eqnarray}
 D(\boldsymbol{\mathcal{X}}_1,\boldsymbol{\mathcal{X}}_2,t) &=& \frac{1}{(2\pi\hbar)^{2\mu}}\int d\boldsymbol{\mathcal{Z}}_1d\boldsymbol{\mathcal{Z}}_2e^{-(i/\hbar)(\boldsymbol{\mathcal{P}}_1\cdot\boldsymbol{\mathcal{Z}}_1+\boldsymbol{\mathcal{P}}_2\cdot\boldsymbol{\mathcal{Z}}_2)}\nonumber\\
&&\times \left\langle \boldsymbol{\mathcal{R}}_2+\frac{\boldsymbol{\mathcal{Z}}_2}{2}\left|e^{-(i/\hbar)\hat{H}t}\hat{\rho}_0\right|\boldsymbol{\mathcal{R}}_1-\frac{\boldsymbol{\mathcal{Z}}_1}{2}\right\rangle\nonumber\\
&& \times \left\langle \boldsymbol{\mathcal{R}}_1+\frac{\boldsymbol{\mathcal{Z}}_1}{2}\left|e^{(i/\hbar)\hat{H}t}\right| \boldsymbol{\mathcal{R}}_2-\frac{\boldsymbol{\mathcal{Z}}_2}{2}\right\rangle,
 \end{eqnarray}
where $\boldsymbol{\mathcal{X}}=(\boldsymbol{\mathcal{R}},\boldsymbol{\mathcal{P}})$. It is interesting to note that $D(\boldsymbol{\mathcal{X}}_1,\boldsymbol{\mathcal{X}}_2,t)$ has the same structure as the spectral density appearing in previous derivations of transport coefficients for mixed quantum-classical systems \cite{Sergi.04.JCP,Kim.05.JCP,Kim.05.JCPa}, with the only difference being that we consider a factorized initial density operator as opposed to a thermal equilibrium state of the total system. Taking into consideration that $\hat{\rho}_0$ and $\hat{H}$ do not commute in general, one can show that $D(\boldsymbol{\mathcal{X}}_1,\boldsymbol{\mathcal{X}}_2,t)$ obeys the following equation of motion (EOM) \cite{Kim.05.JCP} 
 \begin{eqnarray}\label{eq:devo}
 \frac{\partial}{\partial t}D(t) &=& -\frac{i}{\hbar}\left(\hat{H}_W(\boldsymbol{\mathcal{X}}_2)e^{\hbar\widetilde{\Lambda}_2/2i}D(t)\right.\nonumber\\
&& -\left.D(t)e^{\hbar\widetilde{\Lambda}_2/2i}\hat{H}_W(\boldsymbol{\mathcal{X}}_2)\right),
 \end{eqnarray}
 where $\widetilde{\Lambda}_2=\overleftarrow{\nabla}_{\boldsymbol{\mathcal{P}}_2}\overrightarrow{\nabla}_{\boldsymbol{\mathcal{R}}_2}-\overleftarrow{\nabla}_{\boldsymbol{\mathcal{R}}_2}\overrightarrow{\nabla}_{\boldsymbol{\mathcal{P}}_2}$ is the Poisson bracket operator (with the direction of an arrow indicating the direction in which the operator acts).
 
The MGF in Eq.~(\ref{eq:zt_q}) is exact but computationally intractable in general because it involves a fully quantum mechanical treatment of the total system. By taking the quantum-classical limit of Eq.~(\ref{eq:zt_q}), one can obtain an expression that is amenable to numerical simulations. To take this limit, we first note that the full Wigner transform of an operator, $O_W(\boldsymbol{\mathcal{X}})$, may be written as
 \begin{equation}
 O_W(\boldsymbol{\mathcal{X}})~=~\int\,d\boldsymbol{z}e^{(i/\hbar)\boldsymbol{p}\cdot \boldsymbol{z}}\left\langle \boldsymbol{r}-\frac{\boldsymbol{z}}{2}\left|\hat{O}_W(\boldsymbol{X})\right|\boldsymbol{r}+\frac{\boldsymbol{z}}{2}\right\rangle,
 \end{equation}
where $\hat{O}_W(\boldsymbol{X})$ is the partially Wigner-transformed operator. For a quantity that depends only on the variables of the baths, one further has 
 \begin{equation}\label{eq:oo}
O_W(\boldsymbol{\mathcal{X}})~=~O_W(\boldsymbol{X}),
 \end{equation}
i.e., its full Wigner transform is equivalent to its partial Wigner transform. 
This is the case for the exponential functions $(e^{i\sum_v\chi_v H_B^v})_W$ and $(e^{-i\sum_v\chi_v H_B^v})_W$ in Eq.~(\ref{eq:zt_q}). Thus, the MGF in Eq.~(\ref{eq:zt_q}) reduces to
 \begin{eqnarray}\label{eq:zt_q1}
Z(\{\chi_v\},t) &=& \int d\boldsymbol{X}_1d\boldsymbol{X}_2\left(e^{i\sum_v\chi_v H_B^v}\right)_W(\boldsymbol{X}_1)\nonumber\\
&&\times \left(e^{-i\sum_v\chi_v H_B^v}\right)_W(\boldsymbol{X}_2)\overline{D}(\boldsymbol{X}_1,\boldsymbol{X}_2,t),
\end{eqnarray}
where  
\begin{eqnarray}\label{eq:d_full}
&&\overline{D}(\boldsymbol{X}_1,\boldsymbol{X}_2,t) = \int d\boldsymbol{r}_1d\boldsymbol{r}_2d\boldsymbol{p}_1d\boldsymbol{p}_2D(\boldsymbol{\mathcal{X}}_1,\boldsymbol{\mathcal{X}}_2,t)\nonumber\\
&&= \frac{1}{(2\pi\hbar)^{2\mu_B}}\int d\boldsymbol{r}_1d\boldsymbol{Z}_1d\boldsymbol{Z}_2e^{-(i/\hbar)(\boldsymbol{P}_1\cdot \boldsymbol{Z}_1+\boldsymbol{P}_2\cdot \boldsymbol{Z}_2)}\nonumber\\
&&\times \left\langle\boldsymbol{r}_1\left|\left\langle \boldsymbol{R}_1+\frac{\boldsymbol{Z}_1}{2}\left|e^{(i/\hbar)\hat{H}t}\right|\boldsymbol{R}_2-\frac{\boldsymbol{Z}_2}{2}\right\rangle\right.\right.\nonumber\\
&&\times \left.\left.\left\langle \boldsymbol{R}_2+\frac{\boldsymbol{Z}_2}{2}\left|e^{-(i/\hbar)\hat{H}t}\hat{\rho}_0\right|\boldsymbol{R}_1-\frac{\boldsymbol{Z}_1}{2}\right\rangle\right|\boldsymbol{r}_1\right\rangle.
\end{eqnarray}
Similar to Eq.~(\ref{eq:devo}), the EOM for the weight function $\overline{D}(\boldsymbol{X}_1,\boldsymbol{X}_2,t)$ reads
\begin{eqnarray}\label{eq:dbar}
 \frac{\partial}{\partial t}\overline{D}(t) &=& -\frac{i}{\hbar}\left(\hat{H}_W(\boldsymbol{X}_2)e^{\hbar\Lambda_2/2i}\overline{D}(t)\right.\nonumber\\
&& -\left.\overline{D}(t)e^{\hbar\Lambda_2/2i}\hat{H}_W(\boldsymbol{X}_2)\right),
 \end{eqnarray}
 where $\Lambda_2=\overleftarrow{\nabla}_{\boldsymbol{P}_2}\overrightarrow{\nabla}_{\boldsymbol{R}_2}-\overleftarrow{\nabla}_{\boldsymbol{R}_2}\overrightarrow{\nabla}_{\boldsymbol{P}_2}$ is the Poisson bracket operator that acts in the phase space of the heat baths. Up to this point, no approximations were employed, so Eq.~(\ref{eq:zt_q1}) is equivalent to the exact quantum MGF in Eq.~(\ref{eq:zt_total}). However, the difficulties associated with solving Eq.~(\ref{eq:dbar}) are formidable.

The quantum-classical limit of the MGF is then taken by replacing the evolution equation for $\overline{D}(\boldsymbol{X}_1,\boldsymbol{X}_2,t)$ with its quantum-classical limit \cite{Sergi.04.JCP,Kim.05.JCP,Kim.05.JCPa}, i.e., replacing the exponential operator in Eq.~(\ref{eq:dbar}) with its expansion to first order in $\hbar$ \cite{Kapral.99.JCP}:  
 \begin{eqnarray}\label{eq:d}
 \frac{\partial}{\partial t}\overline{D}_{QC}(t) &=& -\frac{i}{\hbar}[\hat{H}_W(\boldsymbol{X}_2),\overline{D}_{QC}(t)]\nonumber\\
 &&+\{\hat{H}_W(\boldsymbol{X}_2),\overline{D}_{QC}(t)\}_a \nonumber \\
 &\equiv& -i\mathcal{L} (\boldsymbol{X}_2)\overline{D}_{QC}(t),
 \end{eqnarray}
where the subscript \textquotedblleft QC\textquotedblright denotes the quantum-classical limit, $\{\hat{H}_W,\cdot\}_a=\frac{1}{2}\{\hat{H}_W,\cdot\}-\frac{1}{2}\{\cdot,\hat{H}_W\}$ is the anti-symmetrized Poisson bracket, and $i\mathcal{L}$ is the quantum-classical Liouville operator.  One can formally solve Eq.~(\ref{eq:d}) to obtain 
\begin{equation} \label{eq:dqc}
\overline{D}_{QC}(\boldsymbol{X}_1,\boldsymbol{X}_2,t)~=~e^{-i\mathcal{L}(\boldsymbol{X}_2)t}\overline{D}(\boldsymbol{X}_1,\boldsymbol{X}_2).
\end{equation}
In the above equation, we note that $\overline{D}(\boldsymbol{X}_1,\boldsymbol{X}_2)$, the zero-time limit of $\overline{D}(\boldsymbol{X}_1,\boldsymbol{X}_2,t)$ in Eq.~(\ref{eq:d_full}), is the initial condition for $\overline{D}_{QC}(t)$. As such, all of the quantum information is retained at the initial time. Thus, the basis-independent quantum-classical MGF is
 \begin{eqnarray}\label{eq:zt_q2}
Z_{QC}(\{\chi_v\},t) &=& \int d\boldsymbol{X}_1d\boldsymbol{X}_2\left(e^{i\sum_v\chi_v H_B^v}\right)_W(\boldsymbol{X}_1)\nonumber\\
&&\times \left(e^{-i\sum_v\chi_v H_B^v}\right)_W(\boldsymbol{X}_2)\nonumber\\
&&\times \overline{D}_{QC}(\boldsymbol{X}_1,\boldsymbol{X}_2,t).
\end{eqnarray}
Noting that the quantum-classical Liouville operator in Eq.~(\ref{eq:dqc}) depends only on $\boldsymbol{X}_2$, one can move the action of the evolution operator onto the term $\left(e^{-i\sum_v\chi_v H_B^v}\right)_W(\boldsymbol{X}_2)$ \cite{Kim.05.JCP} to obtain an equivalent expression for the MGF
 \begin{eqnarray}\label{eq:zt_q3}
Z_{QC}(\{\chi_v\},t) &=& \int d\boldsymbol{X}_1d\boldsymbol{X}_2\left(e^{i\sum_v\chi_v H_B^v}\right)_W(\boldsymbol{X}_1)\nonumber\\
&&\times \left(e^{-i\sum_v\chi_v H_B^v}\right)_W(\boldsymbol{X}_2,t)\nonumber\\
&&\times \overline{D}(\boldsymbol{X}_1,\boldsymbol{X}_2).
\end{eqnarray}
This equation serves as a convenient starting point for computations. 

To evaluate the quantum-classical MGF in Eq.~(\ref{eq:zt_q3}), one must insert complete sets of basis states $\{|\alpha\rangle\}$ that span the Hilbert space of the quantum subsystem
\begin{eqnarray}\label{eq:zt_final}
Z_{QC}(\{\chi_v\},t) 
&=& \sum_{\alpha_1\alpha_2\alpha_2^{\prime}}\int d\boldsymbol{X}_1d\boldsymbol{X}_2\left(e^{i\sum_v\chi_v H_B^v}\right)_W(\boldsymbol{X}_1)\nonumber\\
&&\times \left(e^{-i\sum_v\chi_v H_B^v}\right)_W^{\alpha_2\alpha_2^{\prime}}(\boldsymbol{X}_2,t)\nonumber\\
&&\times \overline{D}^{\alpha_1\alpha_2\alpha_2^{\prime}}(\boldsymbol{X}_1,\boldsymbol{X}_2),
\end{eqnarray}
where
\begin{eqnarray}\label{eq:d_matrix}
&&\overline{D}^{\alpha_1\alpha_2\alpha_2^{\prime}}(\boldsymbol{X}_1,\boldsymbol{X}_2)=\frac{1}{(2\pi\hbar)^{2\mu_B}}\int d\boldsymbol{Z}_1d\boldsymbol{Z}_2\nonumber\\
&&\times e^{-i/\hbar(\boldsymbol{P}_1\cdot \boldsymbol{Z}_1+\boldsymbol{P}_2\cdot \boldsymbol{Z}_2)}\nonumber\\
&&\times \left\langle\alpha_1\left|\left\langle \boldsymbol{R}_1+\boldsymbol{Z}_1/2\left|\boldsymbol{R}_2-\boldsymbol{Z}_2/2\right\rangle\right|\alpha_2\right\rangle\right.\nonumber\\
&& \times \left\langle\alpha_2^{\prime}\left|\left\langle \boldsymbol{R}_2+\boldsymbol{Z}_2/2\left|\hat{\rho}_0\right|\boldsymbol{R}_1-\boldsymbol{Z}_1/2\right\rangle\right|\alpha_1\right\rangle.
\end{eqnarray}
It should be noted that the MGF in Eq. (\ref{eq:zt_final}) has a similar structure to that of a quantum correlation function in the quantum-classical limit \cite{Sergi.04.JCP,Kim.05.JCP,Kim.05.JCPa}. 

\subsection{Average heat and heat current}\label{subsec:1}
In this subsection, we show how one can work out the expected expressions for the average heat and heat current from the quantum-classical MGF in Eq.~(\ref{eq:zt_final}). From the formal expression in Eq.~(\ref{eq:mq}), the average heat is given by 
\begin{eqnarray}\label{eq:q1}
\langle Q_v(t)\rangle
&=& \sum_{\alpha_1\alpha_2}\int d\boldsymbol{X}_1d\boldsymbol{X}_2 \left(H_{B,W}^v\right)(\boldsymbol{X}_1)\nonumber\\
&&\times \overline{D}^{\alpha_1\alpha_2\alpha_2}(\boldsymbol{X}_1,\boldsymbol{X}_2)\nonumber\\
&&- \sum_{\alpha_1\alpha_2\alpha_2^{\prime}}\int d\boldsymbol{X}_1d\boldsymbol{X}_2\left(H_{B,W}^v\right)^{\alpha_2\alpha_2^{\prime}}(\boldsymbol{X}_2,t)\nonumber\\
&&\times \overline{D}^{\alpha_1\alpha_2\alpha_2^{\prime}}(\boldsymbol{X}_1,\boldsymbol{X}_2).
\end{eqnarray}
The first term on the right-hand-side (RHS) of Eq.~(\ref{eq:q1}) may be simplified by first summing over $\alpha_2$ (using the completeness of the basis) and integrating over $\boldsymbol{P}_2$, which results in the delta function $\delta(\boldsymbol{Z}_2)$. The integrals with respect to $\boldsymbol{Z}_2$ and $\boldsymbol{R}_2$ may then be evaluated to yield
\begin{equation}
\sum_{\alpha_1}\int d\boldsymbol{X}_1H_{B,W}^v(\boldsymbol{X}_1)\rho_W^{\alpha_1\alpha_1}(\boldsymbol{X}_1,0),
\end{equation}
where we have used the fact that $\frac{1}{(2\pi\hbar)^{\mu_B}}\int d\boldsymbol{Z}_1e^{-\frac{i}{\hbar}\boldsymbol{P}_1\cdot \boldsymbol{Z}_1}\langle \boldsymbol{R}_1+\boldsymbol{Z}_1/2|\hat{\rho}_0|\boldsymbol{R}_1-\boldsymbol{Z}_1/2\rangle$ is the partially Wigner-transformed initial density matrix $\hat{\rho}_W(0)$. Similarly, for the second term on the RHS of Eq.~(\ref{eq:q1}), we have
\begin{equation}
\sum_{\alpha_2\alpha_2^{\prime}}\int d\boldsymbol{X}_2\left(H_{B,W}^v\right)^{\alpha_2\alpha_2^{\prime}}(\boldsymbol{X}_2,t) \rho_W^{\alpha_2\alpha_2^{\prime}}(\boldsymbol{X}_2,0).
\end{equation}
Summing these two terms together leads to the expression for the average heat in Eq.~(\ref{eq:heat}). By taking the time derivative of this expression, one obtains the expression in Eq.~(\ref{eq:hc}) for the heat current from the $v$th bath to the quantum subsystem.  However, the time evolution in both expressions is now dictated by the quantum-classical Liouville operator.  

\section{Fluctuation symmetry in the quantum-classical limit}\label{sec:4}
It has been previously shown that fully quantum composite systems exhibiting microscopic reversibility obey the following SSFS \cite{Nicolin.11.JCP} (in appendix \ref{a:1}, we provide a proof of this from a closed system point of view) 
\begin{equation}\label{eq:gc}
Z_{ss}(\{\chi_v\},t)~=~Z_{ss}(\{i\beta_v-\chi_v\},t),
\end{equation}
where $Z_{ss}(\{\chi_v\},t)\equiv\lim\limits_{t\to\infty}Z(\{\chi_v\},t)$. Physically speaking, the SSFS determines the heat fluctuations at steady state.  If one now considers a two-heat bath (left and right) setup, the two counting fields $\chi_L$ and $\chi_R$ simply measure the same amount of energy in the steady state. Thus, if one introduces a new counting field $\chi=\chi_R-\chi_L$ (for the case in which the left bath has a higher temperature than the right bath), then the well-known heat exchange fluctuation symmetry is recovered from the SSFS above \cite{Esposito.09.RMP,Campisi.11.RMP}
\begin{equation}
Z_{ss}(\chi,t)~=~Z_{ss}(i\Delta\beta-\chi,t),
\end{equation}
where $\Delta\beta=\beta_R-\beta_L$ is the thermodynamic affinity associated with the steady state heat current. 

In the case of an arbitrary quantum subsystem bilinearly coupled to harmonic baths, the SSFS in Eq.~(\ref{eq:gc}) should exactly hold for the MGF $Z_{QC}(\vec{\chi},t)$ because the QCLE yields the exact quantum dynamics. On the other hand, for systems that go beyond the scope of bilinear interactions and harmonic environments, the approximations inherent to QCLE dynamics may alter the behavior of the heat fluctuations at steady state. It is therefore important to determine to what extent the SSFS holds in such systems.  The MGF in Eq.~(\ref{eq:zt_q1}) is equivalent to the exact quantum MGF and therefore its long-time limit satisfies the SSFS in Eq.~(\ref{eq:gc}).  However, under QCLE dynamics, one can show that the long-time limit of the quantum-classical weight function in Eq.~(\ref{eq:zt_q2}) has the following approximate form (see appendix \ref{a:4} for details)
\begin{equation}\label{eq:d_compare}
\overline{D}_{QC,ss}(\boldsymbol{X}_1,\boldsymbol{X}_2,t) =\overline{D}_{ss}(\boldsymbol{X}_1,\boldsymbol{X}_2,t)+\mathcal{O}_t(\hbar^2),
\end{equation}
where $\overline{D}_{ss}(\boldsymbol{X}_1,\boldsymbol{X}_2,t)\equiv\lim\limits_{t\to\infty}\overline{D}(\boldsymbol{X}_1,\boldsymbol{X}_2,t)$ and $\overline{D}_{QC,ss}(\boldsymbol{X}_1,\boldsymbol{X}_2,t)\equiv\lim\limits_{t\to\infty}\overline{D}_{QC}(\boldsymbol{X}_1,\boldsymbol{X}_2,t)$, and the subscript in $\mathcal{O}_t(\hbar^2)$ indicates that the correction term is time-dependent in the long-time limit. Given the above relation, it follows that the SSFS holds only up to order $\hbar$ under QCLE dynamics (see appendix \ref{a:6} for details), i.e., 
\begin{eqnarray}
&&Z_{QC,ss}(\{\chi_v\},t)=Z_{QC,ss}(\{i\beta_v-\chi_v\},t)+t\cdot\mathcal{O}(\hbar^2),\nonumber\\
&&\mathcal{S}_{QC}(\{\chi_v\})=\mathcal{S}_{QC}(\{i\beta_v-\chi_v\})+\mathcal{O}(\hbar^2),\label{eq:cumu}
\end{eqnarray}
where $Z_{QC,ss}(\{\chi_v\},t)\equiv\lim\limits_{t\to\infty}Z_{QC}(\{\chi_v\},t)$ and $\mathcal{S}_{QC}(\{\chi_v\})\equiv\lim\limits_{t\to\infty}\frac{1}{t}\ln Z_{QC}(\{\chi_v\},t)$ is the quantum-classical scaled cumulant generating function of the heat current.  It should be noted that $\mathcal{S}_{QC}(\{\chi_v\})$ is time-independent because we are considering situations where the cumulants of heat grow linearly with time.  This is usually the case when measuring the statistics of quantities associated with nonequilibrium energy fluxes \cite{Esposito.09.RMP} (anomalous heat statistics have been observed in exceptional cases \cite{Esposito.08.PRE}).

To understand the physical implications of not strictly satisfying the SSFS, we focus on systems with two heat baths
such that the scaled cumulant generating function in Eq.~(\ref{eq:cumu}) reduces to
\begin{equation}\label{eq:cumu_2}
\mathcal{S}_{QC}(\chi)=\mathcal{S}_{QC}(i\Delta\beta-\chi)+\mathcal{O}(\hbar^2).
\end{equation}
By introducing the heat transport coefficients 
\begin{equation}\label{eq:l_def}
L_m^n(\Delta\beta)~\equiv~\left.\frac{\partial^{n+m}}{\partial(i\chi)^n\partial (\Delta\beta)^m}\mathcal{S}_{QC}(\chi)\right|_{\chi=0},
\end{equation}
one can obtain the following Saito-Utsumi (SU) relations \cite{Saito.08.PRB} in the quantum-classical limit 
\begin{equation}\label{eq:su}
L_m^n(\Delta\beta)=\sum_{j=0}^m\left(
\begin{array}{c}
m\\
j
\end{array}
\right)(-1)^{n+j}L_{m-j}^{n+j}(\Delta\beta)+\mathcal{O}(\hbar^2).
\end{equation}
If we now consider the case with $n=0$ and $m=2$, we find that
\begin{equation}\label{eq:fd}
L_1^1(\Delta\beta)~=~\frac{1}{2}L_0^2(\Delta\beta)+\mathcal{O}(\hbar^2),
\end{equation}
where we have used the fact that $L_2^0(\Delta\beta)=0$ due to the normalization condition on the density matrix \cite{Saito.08.PRB}.  In the above equation, the left-hand-side (LHS) equals $\frac{\partial}{\partial \Delta\beta}\langle J\rangle_{ss}$ with $\langle J \rangle_{ss}$ the steady state heat current from the hot bath to the cold one and the RHS is the variance of the steady state heat current. In the linear response regime, the LHS of Eq.~(\ref{eq:fd}) is proportional to the heat conductance. Thus, Eq.~(\ref{eq:fd}) reveals that the fluctuation-dissipation relation is satisfied up to order $\hbar$ in the quantum-classical limit.

We conclude this section by noting that, although the SSFS and fluctuation dissipation theorem are not strictly preserved under QCLE dynamics, the quantum-classical approximation becomes more accurate in the limit that the bath DOF are much heavier than the subsystem DOF and/or under high temperature conditions.

\section{Nonequilibrium spin-boson model}\label{sec:5}

\subsection{Model}
To illustrate the formalism, we consider the NESB model, a prototypical model in the study of quantum energy transfer at the nanoscale \cite{Boudjada.14.JPCA}. This model consists of an unbiased two-level subsystem in contact with two bosonic heat baths at different temperatures. 
The QCLE dynamics of the NESB model is dictated by the following Weyl-ordered, partially Wigner-transformed Hamiltonian
\begin{eqnarray}
\hat{H}_{W} &=& -\hbar\Delta \hat{\sigma}_x+\frac{1}{2}\sum_{v=L,R}\sum_{j=1}^{N_v}\left(P_{j,v}^2+\omega_{j,v}^2R_{j,v}^2\right.\nonumber\\
&&\left.-C_{j,v}R_{j,v}\hat{\sigma}_z-C_{j,v}\hat{\sigma}_zR_{j,v}\right),
\end{eqnarray}
where $\hat{\sigma}_{x/z}$ are the Pauli spin matrices, $\Delta$ is the tunneling frequency between the two states, and $C_{j,v}$ is the coupling coefficient between the spin and the $j$th harmonic oscillator in the $v$th heat bath. The bilinear coupling between the subsystem and $v$th heat bath is characterized by an Ohmic spectral density with an exponential cutoff, namely $I_v(\omega)=\frac{\xi_v}{2}\pi\omega e^{-\omega/\omega_{c,v}}$ with $\xi_v$ the Kondo parameter characterizing the subsystem-bath coupling strength and $\omega_{c,v}$ the cutoff frequency.  In our simulations, we use dimensionless variables and parameters with time scaled by $\omega_c$.

The initial state of the system is chosen to be the product state $\hat{\rho}_{W}(0)=\hat{\rho}_S(0)\rho_{B,W}(0)$, where $\hat{\rho}_S(0)=|+\rangle\langle +|$ ($|+\rangle$ is the spin-up state of $\hat{\sigma}_z$) and $\rho_{B,W}(0)=\prod_v\rho_{B,W}^v(0)$ with 
\begin{eqnarray}\label{eq:rhobw}
\rho_{B,W}^v(0) &=& \prod_{j=1}^{N_v}\frac{\tanh(\hbar\beta_v\omega_{j,v}/2)}{\pi}\exp\left[-\frac{2\tanh(\hbar\beta_v\omega_{j,v}/2)}{\hbar\omega_{j,v}}\right.\nonumber\\
&&\left.\times \left(\frac{P_{j,v}^2}{2}+\frac{\omega_{j,v}^2R_{j,v}^2}{2}\right)\right],
\end{eqnarray}
the partially Wigner-transformed canonical distribution.

Sampling from the weight function $\overline{D}^{\alpha_1\alpha_2\alpha_2^{\prime}}(\boldsymbol{X}_1,\boldsymbol{X}_2)$ in Eq.~(\ref{eq:d_matrix}) remains
a challenging numerical task, so here we focus on simulations of the average heat and heat current.  In this case, $\overline{D}^{\alpha_1\alpha_2\alpha_2^{\prime}}(\boldsymbol{X}_1,\boldsymbol{X}_2)$ reduces to the initial total density matrix (see Sec.~\ref{subsec:1}), which
can be readily sampled from.  The expression for the average heat transferred through the system (obtained from Eq.~(\ref{eq:heat})) is
\begin{eqnarray}\label{eq:h_SB}
\langle Q_v(t)\rangle &=& \sum_{\alpha\alpha^{\prime}}\int d\boldsymbol{X}(0)\rho_{B,W}(\boldsymbol{X}(0))\rho_S^{\alpha^{\prime}\alpha}(0)\nonumber\\
&&\times \sum_{j=1}^{N_v} \left[\frac{P_{j,v}^2(0)\delta_{\alpha\alpha^{\prime}}-(P_{j,v}^2(t))^{\alpha\alpha^{\prime}}}{2}\right.\nonumber\\
&&\left.+\omega_{j,v}^2\frac{R_{j,v}^2(0)\delta_{\alpha\alpha^{\prime}}-(R_{j,v}^2(t))^{\alpha\alpha^{\prime}}}{2}\right],
\end{eqnarray}
where, for example, $\left(P_{j,v}^2(t)\right)^{\alpha\alpha^{\prime}}=\sum_{\beta}P_{j,v}^{\alpha\beta}(t)P_{j,v}^{\beta\alpha^{\prime}}(t)$. From this expression, we see that the time-dependent heat is determined by the time dependence of the matrix elements $(\boldsymbol{P}_{j,v}^2)^{\alpha\alpha^{\prime}}(t)$ and $(\boldsymbol{R}_{j,v}^2)^{\alpha\alpha^{\prime}}(t)$ originating from the bath Hamiltonian $H_{B,W}^v$.  It should be noted that $P_{j,v}^{\alpha\alpha^{\prime}}(t)\neq P_{j,v}(t)\delta_{\alpha\alpha^{\prime}}$ because its time evolution depends on the subsystem's operators due to the subsystem-bath coupling.  The heat current $\langle J_v(t)\rangle$ is defined as the negative of the time derivative of $\langle Q_v(t)\rangle$:
\begin{equation}\label{eq:hc_SB}
\langle J_v(t)\rangle~=~-\frac{d}{dt}\langle Q_v(t)\rangle.
\end{equation}
In our simulations, the heat current is obtained by simply calculating the derivative of $\langle Q_v(t)\rangle$ at each molecular dynamics (MD) time step.

\subsection{Numerical simulations}
\subsubsection{DECIDE solution of QCLE}
In order to evaluate the expressions for the average heat and heat current, we used a recently developed approximate solution of the QCLE known as the DECIDE (Deterministic Evolution of Coordinates with Initial Decoupled Equations) method \cite{Liu.18.NULL}. Instead of propagating the observables directly as in the previous QCLE-based methods, DECIDE evolves the coordinates corresponding to the subsystem and bath (viz., $\boldsymbol{\hat{x}}(t)$ and $\boldsymbol{X}(t)$, respectively) according to the following set of equations of motion (EOMs)
\begin{eqnarray}\label{eq:eom}
\frac{d}{dt}\boldsymbol{\hat{x}}(t) &=& \frac{i}{\hbar}\left([\hat{H}_W,\boldsymbol{\hat{x}}]\right)(t),\nonumber\\
\frac{d}{dt}\boldsymbol{X}(t) &=&-\left(\{\hat{H}_W,\boldsymbol{X}\}_a\right)(t).
\end{eqnarray}
In the above equations, the time arguments are placed outside of their respective brackets to indicate that one should first evaluate the commutator and Poisson brackets with respect to the initial bath coordinates and then apply the time dependence to the coordinates in the resulting expressions.  
  
As the DECIDE algorithm provides an approximate solution of the QCLE, it is worthwhile to discuss the core approximations that enter into the method.  To arrive at Eq.~(\ref{eq:eom}), one starts with the partially Wigner-transformed (with respect to the initial bath coordinates) quantum Heisenberg equations for $\boldsymbol{\hat{x}}(t)$ and $\boldsymbol{\hat{X}}(t)$, and then truncates them by applying the following approximation for an arbitrary time-dependent operator $(\hat{B}(\boldsymbol{\hat{x}}(t),\boldsymbol{\hat{X}}(t)))_W\equiv(e^{i\hat{\mathcal{K}}t}\hat{B}(\boldsymbol{\hat{x}},\boldsymbol{\hat{X}}))_W$
\begin{eqnarray}\label{eq:aaa}
(\hat{B}(\boldsymbol{\hat{x}}(t),\boldsymbol{\hat{X}}(t)))_W &=&
(e^{i\hat{\mathcal{K}}t})_We^{\hbar\Lambda/2i}\hat{B}_W(\boldsymbol{\hat{x}},\boldsymbol{X}) \nonumber \\ &\approx& e^{i\mathcal{L}t}\hat{B}_W(\boldsymbol{\hat{x}},\boldsymbol{X}) \nonumber \\ 
&\equiv& (\hat{B}_W(\boldsymbol{\hat{x}},\boldsymbol{X}))(t),
\end{eqnarray} 
where $\mathcal{\hat{K}}$ is the quantum Liouville operator and $\Lambda$ is the Poisson bracket operator.  To arrive at the second line of this equation, the quantum Liouville operator is replaced with the quantum-classical Liouville operator $i\mathcal{L}$ and only zeroth-order terms in $\hbar$ in the Moyal product expansion are retained.  For the full details of the derivation of Eq.~(\ref{eq:eom}), we refer the readers to Ref.~\cite{Liu.18.NULL}. 

The replacement of the quantum Liouville operator with the quantum-classical Liouville operator in Eq.~(\ref{eq:aaa}) is exact if one considers harmonic environments and bilinear subsystem-bath interactions. However, by neglecting the higher order terms in $\hbar$ in the Moyal product, one may underestimate the back-action from the heat baths to the subsystem even in cases with harmonic environments and bilinear subsystem-bath interactions. To illustrate this, we focus on the first-order correction term to Eq.~(\ref{eq:aaa}), namely $e^{i\mathcal{L}t}\left(\hbar\Lambda/2i\right)\hat{B}_W(\boldsymbol{\hat{x}},\boldsymbol{X})=-i\hbar\{e^{i\mathcal{L}t},\hat{B}_W(\boldsymbol{\hat{x}},\boldsymbol{X})\}$. For demonstration purposes, let us assume that $\hat{B}_W(\boldsymbol{\hat{x}},\boldsymbol{X})\propto \hat{x}\boldsymbol{R}$, which arises from a bilinear subsystem-bath interaction. In this case, the first-order correction term is
\begin{equation}
-i\hbar\left(\frac{\partial}{\partial \boldsymbol{P}}e^{i\mathcal{L}t}\right)\hat{x}=-i\hbar \frac{\partial}{\partial \boldsymbol{P}}\left(e^{i\mathcal{L}t}\hat{x}\right)=-i\hbar\frac{\partial}{\partial \boldsymbol{P}}\hat{x}(t).
\end{equation}
To evaluate the derivative of $\hat{x}(t)$ with respect to the initial momenta, one must know the complete history of the dynamics from the initial time to time $t$. Thus, if the subsystem dynamics is highly non-Markovian, such correction terms cannot be ignored.

In light of its inherent approximations, the DECIDE solution can give rise to inaccurate results in the long-time limit in parameter regimes where non-Markovian effects are pronounced. We note that strong memory effects can be induced by strong subsystem-bath coupling, slow heat baths characterized by $\omega_c\ll\Delta$, and very low temperatures $k_BT_v\ll\hbar \Delta$.  Therefore, DECIDE should be used with caution in such regimes. However, in regimes with weak memory effects, the contributions to the dynamics from the dropped terms in the EOMs for $\boldsymbol{\hat{x}}$ and $\boldsymbol{X}$ are negligible and DECIDE is expected to perform very well (as seen in Ref.~\cite{Liu.18.NULL} and as will be shown below). 

For the NESB model, the generalized coordinates of the spin subsystem are taken to be the Pauli matrices $(i.e., \boldsymbol{\hat{x}}=(\hat{\sigma}_x,\hat{\sigma}_y,\hat{\sigma}_z))$.  Before solving Eq.~(\ref{eq:eom}), one must cast the EOMs in an arbitrary basis $\{|\alpha\rangle\}$ that spans the $2\times2$ Hilbert space of the two-level subsystem, namely
\begin{eqnarray}\label{eq:eom_sb}
\dot{\sigma}_x^{\alpha\alpha^{\prime}}(t) &=& \frac{1}{\hbar}\sum_v\sum_{j=1}^{N_v}C_{j,v}[R_{j,v}(t)\hat{\sigma}_y(t)+\hat{\sigma}_y(t)R_{j,v}(t)]^{\alpha\alpha^{\prime}},\nonumber\\
\dot{\sigma}_y^{\alpha\alpha^{\prime}}(t) &=& 2\Delta\sigma_z^{\alpha\alpha^{\prime}}(t)-\frac{1}{\hbar}\sum_v\sum_{j=1}^{N_v}C_{j,v}[R_{j,v}(t)\hat{\sigma}_x(t)\nonumber\\
&&+\hat{\sigma}_x(t)R_{j,v}(t)]^{\alpha\alpha^{\prime}},\nonumber\\
\dot{\sigma}_z^{\alpha\alpha^{\prime}}(t) &=& -2\Delta\sigma_y^{\alpha\alpha^{\prime}}(t),\nonumber\\
\dot{R}_{j,v}^{\alpha\alpha^{\prime}}(t) &=& P_{j,v}^{\alpha\alpha^{\prime}}(t),\nonumber\\
\dot{P}_{j,v}^{\alpha\alpha^{\prime}}(t) &=& -\omega_{j,v}^2R_{j,v}^{\alpha\alpha^{\prime}}(t)+C_{j,v}\sigma_z^{\alpha\alpha^{\prime}}(t),
\end{eqnarray}
where the dot denotes a time derivative.  In total, there are $4\times(3+2N)$ (with $N=N_L+N_R$) coupled first-order differential equations (FODEs) for the matrix elements $(\sigma_x^{\{\alpha\alpha^{\prime}\}},\sigma_y^{\{\alpha\alpha^{\prime}\}},\sigma_z^{\{\alpha\alpha^{\prime}\}},\boldsymbol{X}^{\{\alpha\alpha^{\prime}\}})$, where $\{\alpha\alpha^{\prime}\}$ denotes all the combinations of basis indices. We remark that the superscript in $\boldsymbol{X}^{\alpha\alpha^{\prime}}(t)$ serves as a label to distinguish the various matrix elements that arise due to the subsystem-bath coupling.

\subsubsection{Simulation details}
To solve the FODEs in Eq.~(\ref{eq:eom_sb}), we must first specify the nature of the $\{|\alpha\rangle\}$  basis set. In this work, we consider two basis sets that are frequently used in studies of the spin-boson model.

The first basis set is a subsystem basis, consisting of the eigenstates of $\hat{\sigma}_z$, i.e., $\{|\alpha\rangle\}=\{|+\rangle,|-\rangle\}$.  In this basis, the initial values of the matrix elements of the subsystem coordinates are $\sigma_x^{+-}(0)=\sigma_x^{-+}(0)=1$, $\sigma_x^{++}(0)=\sigma_x^{--}(0)=0$, $\sigma_y^{++}(0)=\sigma_y^{--}(0) = 0$, $\sigma_y^{-+}=i, \sigma_y^{+-}=-i$, $\sigma_z^{+-}(0)=\sigma_z^{-+}(0)=0$, $\sigma_z^{++}(0)=1,\sigma_z^{--}(0)=-1$; and the initial values of the matrix elements of the bath coordinates are $\boldsymbol{X}^{\alpha\alpha^{\prime}}(0)=\boldsymbol{X}(0)\delta_{\alpha\alpha^{\prime}}$ (due to the initial product state), with $\boldsymbol{X}(0)$ sampled from Eq.~(\ref{eq:rhobw}). In this basis, the expression for the average transferred heat in Eq.~(\ref{eq:h_SB}) reduces to
\begin{eqnarray}\label{eq:h_SB_sub}
\langle Q_v(t)\rangle &=&\int d\boldsymbol{X}(0)\rho_{B,W}(\boldsymbol{X}(0))\nonumber\\
&&\times \sum_{j=1}^{N_v} \left[\frac{P_{j,v}^2(0)-(P_{j,v}^2(t))^{++}}{2}\right.\nonumber\\
&&\left.+\omega_{j,v}^2\frac{R_{j,v}^2(0)-(R_{j,v}^2(t))^{++}}{2}\right],
\end{eqnarray}
using the fact that $\rho_S^{++}(0)=1$. 

The second basis set is the adiabatic basis $\{|\alpha\rangle\}=\{|1\rangle,|2\rangle\}$, which can be expressed in terms of $|\pm\rangle$ as follows \cite{Kernan.02.JCP}
\begin{eqnarray}\label{eq:ab}
|1\rangle &=& \frac{1+G}{\sqrt{2(1+G^2)}}|+\rangle+\frac{1-G}{\sqrt{2(1+G^2)}}|-\rangle,\nonumber\\
|2\rangle &=& \frac{G-1}{\sqrt{2(1+G^2)}}|+\rangle+\frac{1+G}{\sqrt{2(1+G^2)}}|-\rangle,
\end{eqnarray}
where $G=\frac{1}{\gamma(\boldsymbol{R})}[-\Delta+\sqrt{\Delta^2+\gamma(\boldsymbol{R})^2}]$ with $\gamma(\boldsymbol{R})=-\sum_v\sum_{j=1}^{N_v}C_{j,v}R_{j,v}$. In this basis, the initial conditions for the subsystem coordinates are $\sigma_x^{11}(0)=\frac{1-G^2}{1+G^2}, \sigma_x^{12}(0)=\sigma_x^{21}(0)=\frac{2G}{1+G^2}, \sigma_x^{22}(0)=-\frac{1-G^2}{1+G^2}$, $\sigma_y^{12}(0)=-i,\sigma_y^{21}(0) = i$, $\sigma_y^{11}=\sigma_y^{22}=0$, $\sigma_z^{11}(0)=\frac{2G}{1+G^2},\sigma_z^{12}(0)=\sigma_z^{21}(0)=-\frac{1-G^2}{1+G^2}, \textrm{and}~ \sigma_z^{22}(0)=-\frac{2G}{1+G^2}$ (see Appendix \ref{a:5} for details), where $G$ is determined by the initial bath coordinates $\boldsymbol{R}(0)$; the initial values of the bath coordinates are again $\boldsymbol{X}^{\alpha\alpha^{\prime}}(0)=\boldsymbol{X}(0)\delta_{\alpha\alpha^{\prime}}$ (due to the initial product state), with $\boldsymbol{X}(0)$ sampled from Eq.~(\ref{eq:rhobw}). 
It should be noted that, because the adiabatic basis states $\{|1\rangle,|2\rangle\}$ are only used to set the initial values of the coordinates, one does not need to diagonalize the Hamiltonian matrix on-the-fly, in contrast to surface-hopping approaches. After setting the initial values of the coordinates, one just updates them by integrating Eq.~(\ref{eq:eom_sb}).  Using the adiabatic basis, the density operator corresponding to the initial spin-up state is given by
\begin{eqnarray}
\hat{\rho}_S(0) &=& \frac{(1+G)^2}{2(1+G^2)}|1\rangle\langle 1|+\frac{(1-G)^2}{2(1+G^2)}|2\rangle\langle 2|\nonumber\\
&&- \frac{1-G^2}{2(1+G^2)}(|1\rangle\langle 2|+|2\rangle\langle 1|).
\end{eqnarray}
Since the four matrix elements of $\hat{\rho}_S(0)$ are nonzero, there will be four non-zero components in the expression for the average transferred heat in Eq.~(\ref{eq:h_SB}).

To simulate the Ohmic spectral density with the exponential cutoff, we adopt a discretization scheme \cite{Thompson.99.JCP,Wang.01.JCP} with 
\begin{equation}
C_{j,v}~=~\sqrt{\xi_v\hbar\omega_{0,v}}\omega_{j,v},~~~\omega_{j,v}=-\omega_{c,v}\ln\left(1-j\frac{\omega_{0,v}}{\omega_{c,v}}\right),
\end{equation}
where $j$ runs from 1 to $N_v$, $\omega_{0,v}=\frac{\omega_{c,v}}{N_v}(1-e^{-\omega_{m,v}/\omega_{c,v}})$, and $\omega_{m,v}$ is the maximum frequency of the $v$th heat bath. In our simulations, we take $\omega_{m,L}=\omega_{m,R}=\omega_{m}$ and $\omega_{c,L}=\omega_{c,R}=\omega_c$. Although we employ an Ohmic spectral density in this study, it should be emphasized that this approach, just like any other mixed quantum-classical dynamics method, can handle arbitrary bath spectral densities.

Finally, to integrate Eq.~(\ref{eq:eom_sb}), we adopt the standard fourth-order Runge-Kutta scheme \cite{Dormand.80.JCAM}. Noting that $(\boldsymbol{P}_v^2)^{\alpha\alpha^{\prime}}(t)=\sum_{\beta}\boldsymbol{P}_v^{\alpha\beta}(t)\boldsymbol{P}_v^{\beta\alpha^{\prime}}(t)$ and $(\boldsymbol{R}_v^2)^{\alpha\alpha^{\prime}}(t)=\sum_{\beta}\boldsymbol{R}_v^{\alpha\beta}(t)\boldsymbol{R}_v^{\beta\alpha^{\prime}}(t)$, the time evolution of the heat and heat current can then be constructed in terms of the time-dependent coordinates by averaging over an ensemble of trajectories according to Eqs.~(\ref{eq:h_SB}) and (\ref{eq:hc_SB}). 

\subsubsection{Equilibrium condition}
Before considering a temperature gap between the two baths, it is instructive to first consider the equilibrium case ($\beta_L=\beta_R$) and investigate whether the heat currents of the left and right bath vanish at steady state.  In doing so, we also consider symmetric subsystem-bath couplings (i.e., $\xi_L=\xi_R$) and asymmetric ones (i.e., $\xi_L\neq\xi_R$), because numerical methods may predict vanishing heat currents in the long-time limit in one case and fail in the other. 

\begin{figure}[tbh!]
  \centering
  \includegraphics[width=1\columnwidth]{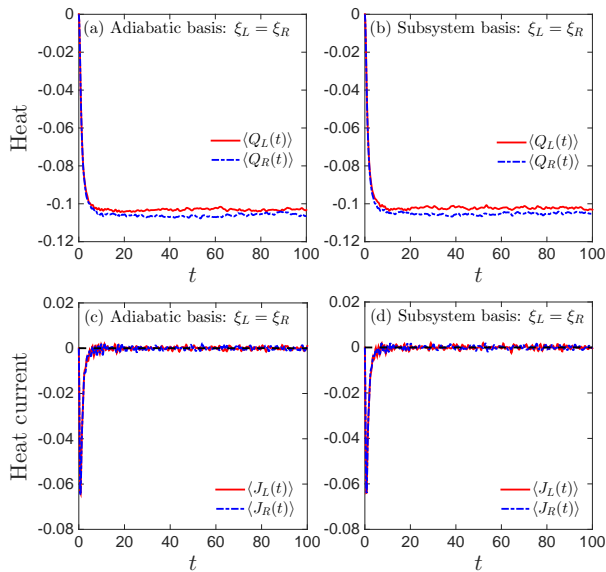}
\caption{The time evolution of the heat and heat current for $\beta_L=\beta_R=0.2$ with symmetric couplings $\xi_L=\xi_R=0.1$. The left and right panels were obtained by using the adiabatic and subsystem bases, respectively. An ensemble of $1\times10^6$ trajectories and a MD time step of $\Delta t=0.02$ were used to obtain converged results. The values of the remaining parameters are $\Delta=0.2$, $\omega_{c}=1$, $\omega_{m}=5$, and $N_L=N_R=150$.}
\label{fig:sb_e1}
\end{figure}

In the symmetric coupling case shown in Fig.~\ref{fig:sb_e1}, we see that both $\langle Q_L(t)\rangle$ and $\langle Q_R(t)\rangle$ become constant (to within numerical error) in the long-time limit, as expected (see Figs.~\ref{fig:sb_e1} (a) and (b) for the results obtained using the adiabatic and subsystem bases, respectively).  In principle, $\langle Q_L(t)\rangle$ and $\langle Q_R(t)\rangle$ should be identical in the symmetric coupling case, but minor deviations between them are observed due to the fact that Eq.~(\ref{eq:eom}) is not exact.   
Nevertheless, the resulting $\langle J_L(t)\rangle$ and $\langle J_R(t)\rangle$ vanish in the long-time limit, implying that the energy conservation condition is satisfied by our simulations. In the transient regime, we find that the left and right bath heat currents are negative, which (according to our sign convention) means that heat is flowing into the baths.  This behaviour has been previously observed and is due to the sudden switch-on of the subsystem-bath couplings at $t=0$ \cite{Cuansing.10.PRB}. 

\begin{figure}[tbh!]
  \centering
  \includegraphics[width=1\columnwidth]{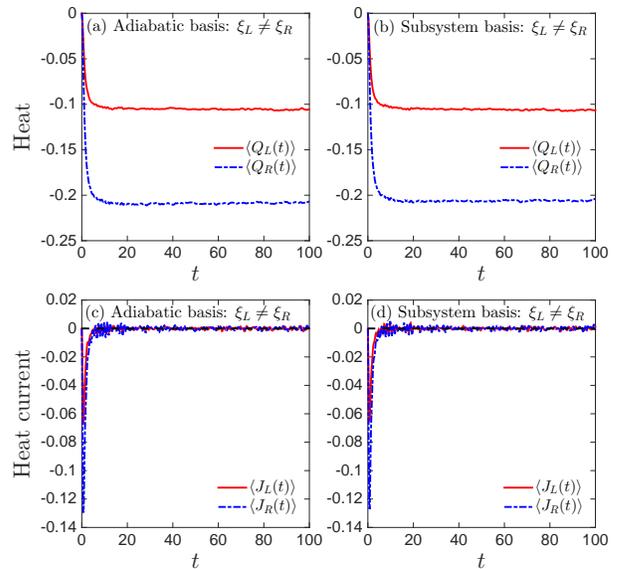}
\caption{The time evolution of the heat and heat currents for $\beta_L=\beta_R=0.2$ with asymmetric couplings $\xi_L=0.1$ and $\xi_R=0.2$. The left and right panels were obtained by using the adiabatic and subsystem bases, respectively. An ensemble of $1\times10^6$ trajectories and a MD time step of $\Delta t=0.02$ were used to obtain converged results. The values of the remaining parameters are $\Delta=0.2$, $\omega_{c}=1$, $\omega_{m}=5$, and $N_L=N_R=150$.}
\label{fig:sb_e2}
\end{figure}

On the other hand, in the asymmetric coupling case shown in Fig.~\ref{fig:sb_e2}, we see that $\langle Q_R(t)\rangle$ (see Figs.~\ref{fig:sb_e2} (a) and (b) for the results obtained using the adiabatic and subsystem bases, respectively) has a larger absolute steady state value due to a larger coupling strength between the subsystem and right heat bath. The ratio between the two steady state heat values is about 2, which is consistent with the ratio of the coupling strengths.
The time dependences of $\langle J_L(t)\rangle$ and $\langle J_R(t)\rangle$ at short times are also different, the latter having a larger drop. However, in the long-time limit, both currents still vanish identically. Furthermore, by comparing the results from the two bases in Figs.~\ref{fig:sb_e1} and \ref{fig:sb_e2}, we find that the results are indeed basis-independent in both the symmetric and asymmetric coupling cases, pointing to the utility of the DECIDE method for simulating heat transfer processes.

\subsubsection{Nonequilibrium condition}
We now consider the nonequilibrium case where the temperatures of the two baths are not equal ($\beta_L\ne\beta_R$). In light of the results of the previous subsection, in this case, we only use the subsystem basis to carry out our calculations and only consider symmetric subsystem-bath couplings ($\xi_L=\xi_R=\xi$). 

\begin{figure}[tbh!]
  \centering
  \includegraphics[width=1\columnwidth]{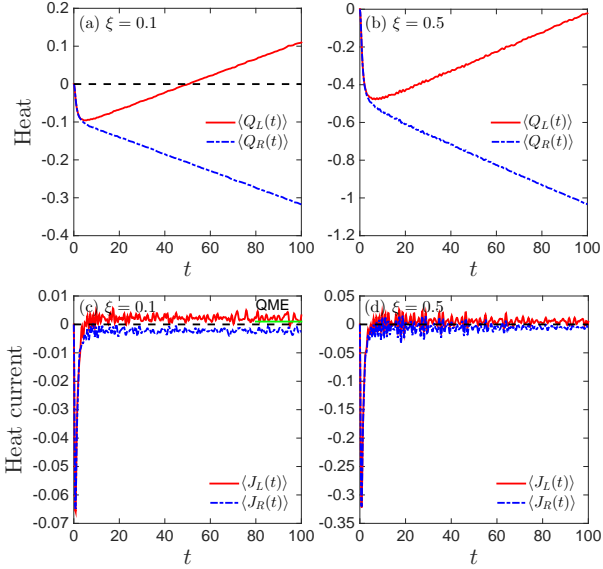}
\caption{The time evolution of the average heat (top) and heat current (bottom) with $\xi=0.1$ (left panels) and $\xi=0.5$ (right panels) at high bath temperatures with $\beta_L=0.1$ and $\beta_R=0.2$. An ensemble of $1\times10^6$ trajectories and a MD time step of $\Delta t=0.02$ were used to obtain converged results. The value of the steady state heat current predicted by the quantum master equation is indicated with a solid green line.\cite{Segal.05.PRL} The values of the remaining parameters are $\Delta=0.2$, $\omega_{c}=1$, $\omega_{m}=5$, and $N_L=N_R=150$.}
\label{fig:sb_n_high}
\end{figure}

\begin{figure}[tbh!]
  \centering
  \includegraphics[width=0.85\columnwidth]{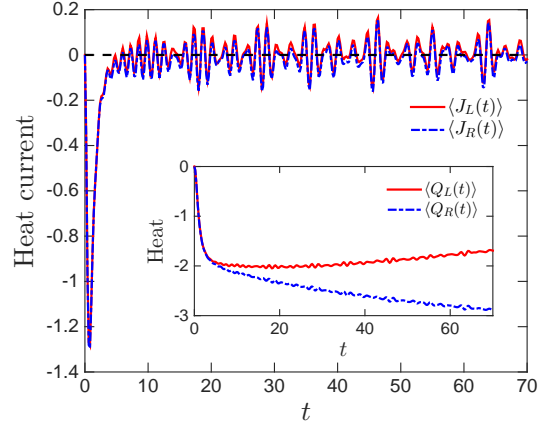}
\caption{The time evolution of the average heat current with $\xi=2$ at high bath temperatures with $\beta_L=0.1$ and $\beta_R=0.2$. An ensemble of $1\times10^6$ trajectories and a MD time step of $\Delta t=0.02$ were used to obtain converged results. The inset shows the dynamics of $\langle Q_L(t)\rangle$ and $\langle Q_R(t)\rangle$. The values of the remaining parameters are $\Delta=0.2$, $\omega_{c}=1$, $\omega_{m}=5$, and $N_L=N_R=150$.}
\label{fig:noncon}
\end{figure}

\begin{figure}[tbh!]
  \centering
  \includegraphics[width=1\columnwidth]{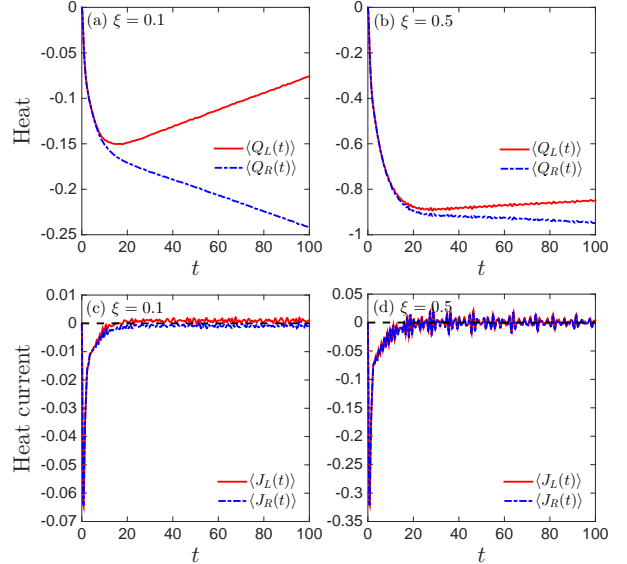}
\caption{The time evolution of the average transferred heat (top) and heat current (bottom) with $\xi=0.1$ (left panels) and $\xi=0.5$ (right panels) at low bath temperatures with $\beta_L=4$ and $\beta_R=6$. An ensemble of $1\times10^6$ trajectories and a MD time step of $\Delta t=0.02$ were used to obtain converged results. The values of the remaining parameters are $\Delta=0.2$, $\omega_{c}=1$, $\omega_{m}=5$, and $N_L=N_R=150$.}
\label{fig:sb_n_low}
\end{figure}

The results for the time-dependent average heat and heat current under different subsystem-bath coupling and temperature conditions are shown in Figs.~\ref{fig:sb_n_high}, \ref{fig:noncon}, and \ref{fig:sb_n_low}. 
We first focus on the results obtained with high bath temperatures in Fig.~\ref{fig:sb_n_high}. From Figs.~\ref{fig:sb_n_high} (a) and (b), we see that, at very short times, both $\langle Q_L(t)\rangle$ and $\langle Q_R(t)\rangle$ are the same because the total system starts from an initial product state and it takes time for the system to adjust to the temperature difference. However, at longer times, $\langle Q_L(t)\rangle$ and $\langle Q_R(t)\rangle$ begin to exhibit differences and ultimately grow linearly with time with opposite slopes, resulting in stationary heat currents.  For $\langle Q_L(t)\rangle$, the slope is positive, so heat is leaving the left (higher temperature) heat bath, while for $\langle Q_R(t)\rangle$, the slope is negative, so heat is entering the right (lower temperature) heat bath.  This behavior was also observed in open quantum linear systems (where a quantum harmonic oscillator is coupled to two harmonic heat baths) by using the nonequilibrium Green's function method \cite{Bijay.12.PRE}. As for the left and right bath heat currents (see Figs.~\ref{fig:sb_n_high} (c) and (d)), their short-time behaviors are similar to those in the equilibrium case. At later times, the currents ultimately plateau with positive and negative values for $\langle J_L(t)\rangle$ and $\langle J_R(t)\rangle$, respectively, i.e., heat is flowing from the hot to the cold bath as expected.  In comparing the left and right panels, we see that, in the strong subsystem-bath coupling case, the oscillations in the heat current are more pronounced (as the back-action from the heat baths becomes stronger) and that the total system takes a longer time to evolve to its steady state.  As a benchmark, in Fig.~\ref{fig:sb_n_high} (c), we provide the value of the steady state heat current predicted by the quantum master equation (QME) \cite{Segal.05.PRL} for the weak subsystem-bath coupling case.  As can be seen, there is a very small deviation between the DECIDE and QME results, which is not surprising as the former is an approximate method and the latter becomes exact in the weak coupling regime at high temperatures. We further note that the behaviors of the heat currents are qualitatively similar to the regularized heat currents obtained by the multilayer multiconfiguration time-dependent Hartree approach in Ref.~\cite{Velizhanin.08.CPL}. 

If we further increase the subsystem-bath coupling strength (from $\xi=0.5$ to $\xi=2$), 
we observe large fluctuations in the heat currents, even though the heat curves are quite smooth (see inset of Fig.~\ref{fig:noncon}).  Due to the large magnitude of the heat current fluctuations, one cannot therefore extract a meaningful steady-state heat current (since the actual steady-state heat current is of order $10^{-3}$ according to the non-interacting blip approximation \cite{Nicolin.11.JCP}). This is expected because the approximation of truncating the exact EOMs for the subsystem and bath coordinates in the DECIDE method deteriorates in the strong coupling regime. Thus, one should be cautious when applying DECIDE to strong subsystem-bath coupling cases.

For heat baths at low temperatures and relatively small subsystem-bath coupling strengths (see Fig.~\ref{fig:sb_n_low}), we find that the transient behaviors of both the transferred heat and heat current are similar to those at high temperatures (see Fig.~\ref{fig:sb_n_high}).  However, we see that the heat current curve in the weaker coupling regime (see Fig.~\ref{fig:sb_n_low} (c)) is smoother than that in the high temperature case (see Fig.~\ref{fig:sb_n_high} (c)) due to the suppression of the thermal noise from the heat baths at lower temperatures. We also notice that the recurrence of the heat current after its initial drop takes longer than in the high temperature case because of the smaller thermodynamic force (resulting from a smaller temperature difference) at the lower temperature.

\section{Summary}\label{sec:6}
In this work, we presented a general formalism for studying nonequilibrium heat transfer processes in mixed quantum-classical systems that combines the FCS and QCLE approaches.  In particular, starting from its exact definition from FCS, we derived a general expression for the MGF of heat in the partial Wigner representation whose dynamics is prescribed by the QCLE.  Using this expression, we obtained explicit expressions for the time-dependent average heat and heat current in a system.  Owing to its mixed quantum-classical nature, this formalism offers a computationally efficient way of studying quantum heat transfer in realistic molecular environments at the nanoscale. 

Since approximations that lead to mixed quantum-classical treatments are expected to alter the behavior of the heat fluctuations at steady state, we further considered to what extent the SSFS holds under QCLE dynamics. We found that the SSFS is preserved up to order $\hbar$ for systems that are beyond the scope of bilinear subsystem-bath interactions and harmonic baths. Using the SU relations, we also showed that a violation of the SSFS is related to a breakdown of the fluctuation-dissipation theorem in linear response regimes. However, if one considers systems in which the bath DOF are much heavier than the subsystem DOF and/or are at high temperatures, the approximations inherent to QCLE dynamics are not expected to significantly affect the SSFS and fluctuation-dissipation theorem.

We demonstrated the performance of this formalism by computing the time-dependent average heat and heat current for the NESB model using the recently developed DECIDE solution of the QCLE.  Under equilibrium conditions (i.e., heat baths at the same temperature), DECIDE yields the expected zero steady state heat currents for both symmetric and asymmetric subsystem-bath couplings.  Under nonequilibrium conditions (i.e., heat baths at different temperatures), DECIDE also yields the expected trends in the average heat and heat current, as compared to previous results obtained with fully quantum methods. Therefore, the present formalism together with the DECIDE method provide a valuable approach for simulating energy transfer processes in open quantum systems out of equilibrium.

Future studies will aim at analyzing the steady state heat current in the NESB model over a wide parameter space using the present method. In particular, it is essential to demonstrate whether the present method can reproduce the turn-over behaviour in the steady state heat current as a function of subsystem-bath coupling strength \cite{Velizhanin.08.CPL}. The generalization of the method to calculate higher order heat fluctuations is also worthwhile. For instance, the noise power of the heat current, obtained from the second order cumulant of the heat, provides rich information beyond what could be inferred from the average heat and heat current \cite{Liu.18.JCP}. We also anticipate applications of the method to multi-level (i.e., beyond two levels) subsystems with more complex environments.  Finally, one could consider applying other mixed quantum-classical and semi-classical methods, such as those previously used in the study of vibrational energy transfer in condensed phases \cite{Shi.03.JPCA,Hanna.08.JPCB,Jain.18.JPCA}, to nonequilibrium heat transfer problems.

\begin{acknowledgments}
J. Liu and G. Hanna acknowledge support from the Natural Sciences and Engineering Research Council of Canada (NSERC). C.-Y. Hsieh acknowledges support from the Singapore-MIT Alliance for Research and Technology (SMART). D. Segal acknowledges support from an NSERC Discovery Grant and the Canada Research Chair program.
\end{acknowledgments}

\appendix
\section{Fluctuation symmetry in the long-time limit}\label{a:1}
In this appendix, we provide an alternative proof to that in Ref.~\cite{Nicolin.11.JCP} that, in quantum composite systems with time-reversal symmetry, the MGF of heat satisfies the symmetry relation $Z_{ss}(\{\chi_v\},t)=Z_{ss}(\{i\beta_v-\chi_v\},t)$ in the long-time limit, where $\beta_v$ is the inverse temperature of $v$th heat bath. 

To start, by using the facts that the trace in Eq.~(\ref{eq:zt_total}) is invariant to cyclic permutation and that $[\hat{\rho}_0, \hat{H}_B^v]=0$, we may re-express the MGF as
\begin{equation}\label{eq:ztd2}
Z(\{\chi_v\},t) ~=~ \mathrm{Tr}[e^{-i\sum_v\chi_v \hat{H}_B^v(t)}e^{i\sum_v\chi_v \hat{H}_B^v}\hat{\rho}_0].
\end{equation}
In the absence of any external driving, we can shift the time arguments in Eq.~(\ref{eq:ztd2}) as follows by noting that $\hat{U}(t,0)=\hat{U}(t/2,0)\hat{U}(t/2,0)$:
\begin{equation}\label{eq:ztd7}
Z(\{\chi_v\},t)~=~\mathrm{Tr}\left[e^{-i\sum_v\chi_v \hat{H}_B^v\left(\frac{t}{2}\right)}e^{i\sum_v\chi_v \hat{H}_B^v\left(-\frac{t}{2}\right)}\hat{\rho}\left(t/2\right)\right],
\end{equation}
with $\hat{\rho}\left(t/2\right)\equiv\hat{U}(t/2,0)\hat{\rho}_0\hat{U}^{\dagger}(t/2,0)$. 
Next, applying the transformation $\chi_v\to i\beta_v-\chi_v$ to Eq. (\ref{eq:ztd7}), we obtain
\begin{eqnarray}\label{eq:ztd6}
&&Z(\{i\beta_v-\chi_v\},t) = \mathrm{Tr}\left[e^{i\sum_v\chi_v \hat{H}_B^v\left(\frac{t}{2}\right)}e^{-i\sum_v\chi_v \hat{H}_B^v\left(-\frac{t}{2}\right)}\right.\nonumber\\
&&\times \left.e^{-\sum_v\beta_v \hat{H}_B^v\left(-\frac{t}{2}\right)}\hat{\rho}\left(t/2\right)e^{\sum_v\beta_v \hat{H}_B^v\left(\frac{t}{2}\right)}\right].
\end{eqnarray}
Since $\hat{\rho}_0$ commutes with the bath Hamiltonians, we can show that 
\begin{eqnarray} \label{eq:ztd8}
&&e^{-\sum_v\beta_v \hat{H}_B^v\left(-\frac{t}{2}\right)}\hat{\rho}\left(t/2\right)e^{\sum_v\beta_v \hat{H}_B^v\left(\frac{t}{2}\right)}\nonumber\\
&& = \hat{\rho}\left(t/2\right)e^{-\sum_v\beta_v \hat{H}_B^v\left(-\frac{t}{2}\right)}e^{\sum_v\beta_v \hat{H}_B^v\left(\frac{t}{2}\right)}.
\end{eqnarray}
In the limit of $t\to\infty$, the RHS of Eq.~(\ref{eq:ztd8}) becomes
\begin{equation} \label{ztd9}
\hat{\rho}_{ss}e^{-\sum_v\beta_v \hat{H}_B^v\left(-\infty\right)}e^{\sum_v\beta_v \hat{H}_B^v\left(\infty\right)}.
\end{equation}
Given the micro-reversibility of the closed system without external driving, one may argue that $\hat{H}_B^v(\infty)$ and $\hat{H}_B^v(-\infty)$ are equal and, therefore, Eq.~(\ref{ztd9}) reduces to $\hat{\rho}_{ss}$ in the long-time limit.  

To complete the proof, we will need the following relations for an arbitrary operator $\hat{B}$:
\begin{equation} \label{eq:trev}
\mathrm{Tr}[\Theta^{\dagger}\hat{B}(t)\Theta\hat{\rho}_{ss}]~=~\mathrm{Tr}[\hat{B}^{\dagger}(-t)\hat{\rho}_{ss}]=\mathrm{Tr}[\hat{B}^{\dagger}(t)\hat{\rho}_{ss}],
\end{equation}
where $\Theta$ is the quantum mechanical time-reversal operator and $\hat{\rho}_{ss}$ is the steady state density operator of the total system. In arriving to the last equality in the above equation, we used the fact that the time evolution operator commutes with $\hat{\rho}_{ss}$. 
Now, similarly to what is done above, we can shift the time arguments in Eq.~(\ref{eq:zt_total}), take the long-time limit, and use Eq.~(\ref{eq:trev}) to yield the following: 
\begin{equation}\label{eq:ztd5}
Z_{ss}(\{\chi_v\},t)~=\mathrm{Tr}\left[e^{i\sum_v\chi_v \hat{H}_B^v\left(\infty\right)}e^{-i\sum_v\chi_v \hat{H}_B^v\left(-\infty\right)}\hat{\rho}_{ss}\right].
\end{equation}
Finally, comparing Eq.~(\ref{eq:ztd5}) with the long-time limit of Eq.~(\ref{eq:ztd6}) (obtained with the aid of Eq.~(\ref{ztd9})), we see that
\begin{equation}
Z_{ss}(\{\chi_v\},t)~=~Z_{ss}(\{i\beta_v-\chi_v\},t).
\end{equation}

\section{Long-time limit of $\overline{D}_{QC}$}\label{a:4}
Given Eqs.~(\ref{eq:dbar}) and (\ref{eq:d}), $\overline{D}$ and $\overline{D}_{QC}$ in the long-time limit should respectively satisfy the following equations:
\begin{eqnarray} \label{eq:ssD}
&&-\frac{i}{\hbar}\left(\hat{H}_W e^{\hbar\Lambda_2/2i}\overline{D}_{ss}-\overline{D}_{ss}e^{\hbar\Lambda_2/2i}\hat{H}_W\right) =\frac{\partial}{\partial t}\overline{D}_{ss},\nonumber\\
&&-\frac{i}{\hbar}\left(\hat{H}_W(1+\frac{\hbar\Lambda_2}{2i})\overline{D}_{QC,ss}-\overline{D}_{QC,ss}(1+\frac{\hbar\Lambda_2}{2i})\hat{H}_W\right)\nonumber\\
&&=\frac{\partial}{\partial t}\overline{D}_{QC,ss}.
\end{eqnarray}
In order to analyze the connection between $\overline{D}_{ss}(t)$ and its approximated form $\overline{D}_{QC,ss}(t)$, we first expand these quantities in power series of $\hbar$ \cite{Nielsen.01.JCP}:
\begin{eqnarray}
\overline{D}_{ss}(t) &=& \sum_{n=0}^{\infty}\hbar^n\overline{D}_{ss}^{(n)}(t),\label{eq:dss}\\
\overline{D}_{QC,ss}(t) &=& \sum_{n=0}^{\infty}\hbar^n\overline{D}_{QC,ss}^{(n)}(t)\label{eq:dqcss}.
\end{eqnarray}
We then substitute these power series back into Eq.~(\ref{eq:ssD}) and group by powers of $\hbar$. For $\overline{D}_{ss}(t)$, this leads to
\begin{eqnarray}\label{eq:dss_order}
(\hbar^{0}~\mathrm{order}):&&0=-i[\hat{H}_W,\overline{D}_{ss}^{(0)}],\nonumber\\
(\hbar^{1}~\mathrm{order}):&&\frac{\partial}{\partial t}\overline{D}_{ss}^{(0)}=-i[\hat{H}_W,\overline{D}_{ss}^{(1)}]-\{\overline{D}_{ss}^{(0)},\hat{H}_W\}_a,\nonumber\\
(\hbar^{2}~\mathrm{order}):&&\frac{\partial}{\partial t}\overline{D}_{ss}^{(1)}=-i[\hat{H}_W,\overline{D}_{ss}^{(2)}]-\{\overline{D}_{ss}^{(1)},\hat{H}_W\}_a\nonumber\\
&&+\frac{i}{8}\hat{H}_W\Lambda_2^2\overline{D}_{ss}^{(0)}-\frac{i}{8}\overline{D}_{ss}^{(0)}\Lambda_2^2\hat{H}_W, 
\end{eqnarray}
and so on. 
For $\overline{D}_{QC,ss}$, this leads to the following recursion relations: 
\\For $\hbar^{0}$ order, 
\begin{equation}\label{eq:dqcss_1}
0~=~-i[\hat{H}_W,\overline{D}_{QC,ss}^{(0)}],
\end{equation}
and for $\hbar^{n}$ order with $n\geqslant 1$,
 \begin{equation}\label{eq:dqcss_n}
\frac{\partial}{\partial t}\overline{D}_{QC,ss}^{(n-1)}~=~-i[\hat{H}_W,\overline{D}_{QC,ss}^{(n)}]-\{\overline{D}_{QC,ss}^{(n-1)},\hat{H}_W\}_a.
 \end{equation}
Comparing Eqs.~(\ref{eq:dqcss_1}) and (\ref{eq:dqcss_n}) with Eq.~(\ref{eq:dss_order}), we find that $\overline{D}_{QC,ss}(t)$ and $\overline{D}_{ss}(t)$ are identical to order $\hbar$, namely, $\overline{D}_{QC,ss}^{(0)}(t)=\overline{D}_{ss}^{(0)}(t)$, $\overline{D}_{QC,ss}^{(1)}(t)=\overline{D}_{ss}^{(1)}(t)$, and $\overline{D}_{QC,ss}^{(n)}(t)\neq\overline{D}_{ss}^{(n)}(t)$ for $n>1$, thereby proving Eq.~(\ref{eq:d_compare}) in the main text.

\section{Fluctuation symmetry in the quantum-classical limit}\label{a:6}
The long-time limits of the quantum and quantum-classical MGFs in Eqs.~(\ref{eq:zt_q1}) and (\ref{eq:zt_q2}), respectively, are
\begin{eqnarray}
&&Z_{ss}(\{\chi_v\},t) =\int d\boldsymbol{X}_1d\boldsymbol{X}_2\left(e^{i\sum_v\chi_v H_B^v}\right)_W(\boldsymbol{X}_1)\nonumber\\
&&\times \left(e^{-i\sum_v\chi_v H_B^v}\right)_W(\boldsymbol{X}_2)\overline{D}_{ss}(\boldsymbol{X}_1,\boldsymbol{X}_2,t),\\
&&Z_{QC,ss}(\{\chi_v\},t)=\int d\boldsymbol{X}_1d\boldsymbol{X}_2\left(e^{i\sum_v\chi_v H_B^v}\right)_W(\boldsymbol{X}_1)\nonumber\\
&&\times \left(e^{-i\sum_v\chi_v H_B^v}\right)_W(\boldsymbol{X}_2)\overline{D}_{QC,ss}(\boldsymbol{X}_1,\boldsymbol{X}_2,t),
\end{eqnarray}
where $Z_{QC,ss}(\{\chi_v\},t)\equiv\lim\limits_{t\to\infty}Z_{QC}(\{\chi_v\},t)$.  We can expand these MGFs in power series of $\hbar$ (as done in Eqs.~(\ref{eq:dss}) and (\ref{eq:dqcss})) to yield 
\begin{eqnarray}
Z_{ss}(\{\chi_v\},t) &=& \sum_{n=0}^{\infty}\hbar^nZ_{ss}^{(n)}(\{\chi_v\},t),\label{eq:zss}\\
Z_{QC,ss}(\{\chi_v\},t) &=& \sum_{n=0}^{\infty}\hbar^nZ_{QC,ss}^{(n)}(\{\chi_v\},t)\label{eq:zqcss}
\end{eqnarray}
with $Z_{ss}^{(n)}(\{\chi_v\},t)$ and $Z_{QC,ss}^{(n)}(\{\chi_v\},t)$ solely determined by $\overline{D}_{ss}^{(n)}(t)$ and $\overline{D}_{QC,ss}^{(n)}(t)$, respectively, e.g., $Z_{ss}^{(2)}(\{\chi_v\},t)$ does not depend on $\overline{D}_{ss}^{(1)}(t)$ because they are associated with different orders of $\hbar$. 
Given the analysis in appendix \ref{a:4} for the weight functions, one can conclude that $Z_{QC,ss}(\{\chi_v\},t)$ and $Z_{ss}(\{\chi_v\},t)$ are also identical to order of $\hbar$, i.e., 
\begin{equation}\label{eq:z_compare}
Z_{QC,ss}(\{\chi_v\},t)~=~Z_{ss}(\{\chi_v\},t)+\mathcal{O}_t(\hbar^2),
\end{equation}
where $\mathcal{O}_t(\hbar^2)$ is time-dependent. Since $Z_{ss}(\{\chi_v\},t)$ preserves the SSFS, it immediately follows that
\begin{equation}\label{eq:qc_ssfs}
Z_{QC,ss}(\{\chi_v\},t)~=~Z_{QC,ss}(\{i\beta_v-\chi_v\},t)+\mathcal{O}_t(\hbar^2).
\end{equation}

Based on Eq.~(\ref{eq:mq}), we can further rewrite the MGFs in terms of the moments of heat as 
\begin{eqnarray}
Z_{ss}(\chi_v,t) &=& \sum_{m=0}^{\infty}\frac{(i\chi_v)^m}{m!}\left\langle Q_v^m(t)\right\rangle_{ss},\label{eq:zss_1}\\
Z_{QC,ss}(\chi_v,t) &=& \sum_{m=0}^{\infty}\frac{(i\chi_v)^m}{m!}\left\langle Q_v^m(t)\right\rangle_{QC,ss}\label{eq:zqcss_1},
\end{eqnarray}
where the other counting fields except for $\chi_v$ are zero. If we now expand the moments of heat in power series of $\hbar$, and compare Eqs.~(\ref{eq:zss}) and (\ref{eq:zqcss}) with Eqs. (\ref{eq:zss_1}) and (\ref{eq:zqcss_1}), respectively, we obtain the following relations 
\begin{eqnarray}
Z_{ss}^{(n)}(\chi_v,t) &=& \sum_{m=0}^{\infty}\frac{(i\chi_v)^m}{m!}\left\langle Q_v^m(t)\right\rangle_{ss}^{(n)},\label{eq:zss_2}\\
Z_{QC,ss}^{(n)}(\chi_v,t) &=& \sum_{m=0}^{\infty}\frac{(i\chi_v)^m}{m!}\left\langle Q_v^m(t)\right\rangle_{QC,ss}^{(n)}\label{eq:zqcss_2}.
\end{eqnarray}
Therefore, according to Eq.~(\ref{eq:z_compare}), we find that the moments of heat $\left\langle Q_v^m(t)\right\rangle_{ss}$ and $\left\langle Q_v^m(t)\right\rangle_{QC,ss}$ are identical to order $\hbar$, i.e., 
\begin{equation}\label{eq:m_compare}
\left\langle Q_v^m(t)\right\rangle_{QC,ss}~=~\left\langle Q_v^m(t)\right\rangle_{ss}+\mathcal{O}_t(\hbar^2).
\end{equation} 

Next, we introduce the long-time limits of the cumulant generating functions (CGFs) of heat $G_{ss}(\{\chi_v\},t)\equiv\ln Z_{ss}(\{\chi_v\},t)$ and $G_{QC,ss}(\{\chi_v\},t)\equiv\ln Z_{QC,ss}(\{\chi_v\},t)$. Given these definitions, one can establish the following relationship between the quantum and quantum-classical cumulants of heat $\left\langle\langle Q_v^m(t)\right\rangle\rangle_{ss}$ and $\left\langle\langle Q_v^m(t)\right\rangle\rangle_{QC,ss}$, respectively:
\begin{equation}\label{eq:c_compare}
\left\langle\langle Q_v^m(t)\right\rangle\rangle_{QC,ss}~=~\left\langle\langle Q_v^m(t)\right\rangle\rangle_{ss}+\mathcal{O}_t(\hbar^2).
\end{equation}
In most cases, the cumulants of heat grow linearly with time \cite{Esposito.09.RMP}, so one may argue that $\mathcal{O}_t(\hbar^2)\sim t\cdot\mathcal{O}(\hbar^2)$ in the above equation and consequently in Eq.~(\ref{eq:qc_ssfs}), thereby proving the first equation in Eq.~(\ref{eq:cumu}). Given this linear time dependence of the cumulants in the long-time limit, one may define time-independent scaled CGFs of the heat current as follows
\begin{equation}
\mathcal{S}_{QC}(\{\chi_v\})\equiv\frac{1}{t}G_{QC,ss}(\{\chi_v\},t),~~~\mathcal{S}(\{\chi_v\})\equiv\frac{1}{t}G_{ss}(\{\chi_v\},t).
\end{equation}
Since the CGFs can also be expanded in terms of the cumulants of heat (in analogy with Eqs.~(\ref{eq:zss_1}) and (\ref{eq:zqcss_1}) for the MGFs), Eq.~(\ref{eq:c_compare}) implies that
\begin{equation}\label{eq:G_compare}
\mathcal{S}_{QC}(\{\chi_v\})~=~\mathcal{S}(\{\chi_v\})+\mathcal{O}(\hbar^2).
\end{equation}
Finally, given the fact that $\mathcal{S}(\{\chi_v\})$ preserves the SSFS, one can recover the second equation in Eq.~(\ref{eq:cumu}).

\section{Pauli matrices in the adiabatic basis}\label{a:5}
Using Eq.~(\ref{eq:ab}), one can express the Pauli matrices in the adiabatic basis as
\begin{eqnarray}
\hat{\sigma}_x &=& \frac{1-G^2}{1+G^2}|1\rangle\langle 1|- \frac{1-G^2}{1+G^2}|2\rangle\langle 2|\nonumber\\
&&+\frac{2G}{1+G^2}(|1\rangle\langle 2|+|2\rangle\langle 1|),\nonumber
\end{eqnarray}
\begin{eqnarray}
\hat{\sigma}_y &=& -i|1\rangle\langle 2|+i|2\rangle\langle 1|,\nonumber\\
\hat{\sigma}_z &=& \frac{2G}{1+G^2}|1\rangle\langle 1|- \frac{2G}{1+G^2}|2\rangle\langle 2|\nonumber\\
&&-\frac{1-G^2}{1+G^2}(|1\rangle\langle 2|+|2\rangle\langle 1|).
\end{eqnarray}
From these expressions, one can determine the initial values of the subsystem coordinates given below Eq.~(\ref{eq:ab}).

\bibliography{a}

\begin{thebibliography}{70}
\expandafter\ifx\csname natexlab\endcsname\relax\def\natexlab#1{#1}\fi
\expandafter\ifx\csname bibnamefont\endcsname\relax
  \def\bibnamefont#1{#1}\fi
\expandafter\ifx\csname bibfnamefont\endcsname\relax
  \def\bibfnamefont#1{#1}\fi
\expandafter\ifx\csname citenamefont\endcsname\relax
  \def\citenamefont#1{#1}\fi
\expandafter\ifx\csname url\endcsname\relax
  \def\url#1{\texttt{#1}}\fi
\expandafter\ifx\csname urlprefix\endcsname\relax\def\urlprefix{URL }\fi
\providecommand{\bibinfo}[2]{#2}
\providecommand{\eprint}[2][]{\url{#2}}

\bibitem[{\citenamefont{Wang et~al.}({2007})\citenamefont{Wang, Carter,
  Lagutchev, Koh, Seong, Cahill, and Dlott}}]{Wang.07.S}
\bibinfo{author}{\bibfnamefont{Z.}~\bibnamefont{Wang}},
  \bibinfo{author}{\bibfnamefont{J.~A.} \bibnamefont{Carter}},
  \bibinfo{author}{\bibfnamefont{A.}~\bibnamefont{Lagutchev}},
  \bibinfo{author}{\bibfnamefont{Y.~K.} \bibnamefont{Koh}},
  \bibinfo{author}{\bibfnamefont{N.-H.} \bibnamefont{Seong}},
  \bibinfo{author}{\bibfnamefont{D.~G.} \bibnamefont{Cahill}},
  \bibnamefont{and} \bibinfo{author}{\bibfnamefont{D.~D.} \bibnamefont{Dlott}},
  \bibinfo{journal}{{Science}} \textbf{\bibinfo{volume}{{317}}},
  \bibinfo{pages}{{787}} (\bibinfo{year}{{2007}}).

\bibitem[{\citenamefont{Schwab et~al.}(2000)\citenamefont{Schwab, Henriksen,
  Worlock, and Roukes}}]{Schwab.00.N}
\bibinfo{author}{\bibfnamefont{K.}~\bibnamefont{Schwab}},
  \bibinfo{author}{\bibfnamefont{E.~A.} \bibnamefont{Henriksen}},
  \bibinfo{author}{\bibfnamefont{J.~M.} \bibnamefont{Worlock}},
  \bibnamefont{and} \bibinfo{author}{\bibfnamefont{M.~L.}
  \bibnamefont{Roukes}}, \bibinfo{journal}{Nature}
  \textbf{\bibinfo{volume}{404}}, \bibinfo{pages}{974} (\bibinfo{year}{2000}).

\bibitem[{\citenamefont{Carter et~al.}(2009)\citenamefont{Carter, Wang, and
  Dlott}}]{Carter.09.ACR}
\bibinfo{author}{\bibfnamefont{J.~A.} \bibnamefont{Carter}},
  \bibinfo{author}{\bibfnamefont{Z.}~\bibnamefont{Wang}}, \bibnamefont{and}
  \bibinfo{author}{\bibfnamefont{D.~D.} \bibnamefont{Dlott}},
  \bibinfo{journal}{Acc. Chem. Res.} \textbf{\bibinfo{volume}{42}},
  \bibinfo{pages}{1343} (\bibinfo{year}{2009}).

\bibitem[{\citenamefont{Losego et~al.}(2012)\citenamefont{Losego, Grady,
  Sottos, Cahill, and Braun}}]{Mark.12.NM}
\bibinfo{author}{\bibfnamefont{M.~D.} \bibnamefont{Losego}},
  \bibinfo{author}{\bibfnamefont{M.~E.} \bibnamefont{Grady}},
  \bibinfo{author}{\bibfnamefont{N.~R.} \bibnamefont{Sottos}},
  \bibinfo{author}{\bibfnamefont{D.~G.} \bibnamefont{Cahill}},
  \bibnamefont{and} \bibinfo{author}{\bibfnamefont{P.~V.} \bibnamefont{Braun}},
  \bibinfo{journal}{Nat. Mater.} \textbf{\bibinfo{volume}{11}},
  \bibinfo{pages}{502} (\bibinfo{year}{2012}).

\bibitem[{\citenamefont{Meier et~al.}(2014)\citenamefont{Meier, Menges,
  Nirmalraj, H\"olscher, Riel, and Gotsmann}}]{Meier.14.PRL}
\bibinfo{author}{\bibfnamefont{T.}~\bibnamefont{Meier}},
  \bibinfo{author}{\bibfnamefont{F.}~\bibnamefont{Menges}},
  \bibinfo{author}{\bibfnamefont{P.}~\bibnamefont{Nirmalraj}},
  \bibinfo{author}{\bibfnamefont{H.}~\bibnamefont{H\"olscher}},
  \bibinfo{author}{\bibfnamefont{H.}~\bibnamefont{Riel}}, \bibnamefont{and}
  \bibinfo{author}{\bibfnamefont{B.}~\bibnamefont{Gotsmann}},
  \bibinfo{journal}{Phys. Rev. Lett.} \textbf{\bibinfo{volume}{113}},
  \bibinfo{pages}{060801} (\bibinfo{year}{2014}).

\bibitem[{\citenamefont{Cui et~al.}(2017)\citenamefont{Cui, Jeong, Hur, Matt,
  Kl{\"o}ckner, Pauly, Nielaba, Cuevas, Meyhofer, and Reddy}}]{Cui.17.S}
\bibinfo{author}{\bibfnamefont{L.}~\bibnamefont{Cui}},
  \bibinfo{author}{\bibfnamefont{W.}~\bibnamefont{Jeong}},
  \bibinfo{author}{\bibfnamefont{S.}~\bibnamefont{Hur}},
  \bibinfo{author}{\bibfnamefont{M.}~\bibnamefont{Matt}},
  \bibinfo{author}{\bibfnamefont{J.~C.} \bibnamefont{Kl{\"o}ckner}},
  \bibinfo{author}{\bibfnamefont{F.}~\bibnamefont{Pauly}},
  \bibinfo{author}{\bibfnamefont{P.}~\bibnamefont{Nielaba}},
  \bibinfo{author}{\bibfnamefont{J.~C.} \bibnamefont{Cuevas}},
  \bibinfo{author}{\bibfnamefont{E.}~\bibnamefont{Meyhofer}}, \bibnamefont{and}
  \bibinfo{author}{\bibfnamefont{P.}~\bibnamefont{Reddy}},
  \bibinfo{journal}{Science} \textbf{\bibinfo{volume}{355}},
  \bibinfo{pages}{1192} (\bibinfo{year}{2017}).

\bibitem[{\citenamefont{Segal and Nitzan}(2005)}]{Segal.05.PRL}
\bibinfo{author}{\bibfnamefont{D.}~\bibnamefont{Segal}} \bibnamefont{and}
  \bibinfo{author}{\bibfnamefont{A.}~\bibnamefont{Nitzan}},
  \bibinfo{journal}{Phys. Rev. Lett.} \textbf{\bibinfo{volume}{94}},
  \bibinfo{pages}{034301} (\bibinfo{year}{2005}).

\bibitem[{\citenamefont{Velizhanin et~al.}(2008)\citenamefont{Velizhanin, Wang,
  and Thoss}}]{Velizhanin.08.CPL}
\bibinfo{author}{\bibfnamefont{K.~A.} \bibnamefont{Velizhanin}},
  \bibinfo{author}{\bibfnamefont{H.}~\bibnamefont{Wang}}, \bibnamefont{and}
  \bibinfo{author}{\bibfnamefont{M.}~\bibnamefont{Thoss}},
  \bibinfo{journal}{Chem. Phys. Lett.} \textbf{\bibinfo{volume}{460}},
  \bibinfo{pages}{325} (\bibinfo{year}{2008}).

\bibitem[{\citenamefont{Ren et~al.}(2010)\citenamefont{Ren, H\"anggi, and
  Li}}]{Ren.10.PRL}
\bibinfo{author}{\bibfnamefont{J.}~\bibnamefont{Ren}},
  \bibinfo{author}{\bibfnamefont{P.}~\bibnamefont{H\"anggi}}, \bibnamefont{and}
  \bibinfo{author}{\bibfnamefont{B.}~\bibnamefont{Li}}, \bibinfo{journal}{Phys.
  Rev. Lett.} \textbf{\bibinfo{volume}{104}}, \bibinfo{pages}{170601}
  (\bibinfo{year}{2010}).

\bibitem[{\citenamefont{Nicolin and
  Segal}(2011{\natexlab{a}})}]{Nicolin.11.JCP}
\bibinfo{author}{\bibfnamefont{L.}~\bibnamefont{Nicolin}} \bibnamefont{and}
  \bibinfo{author}{\bibfnamefont{D.}~\bibnamefont{Segal}}, \bibinfo{journal}{J.
  Chem. Phys.} \textbf{\bibinfo{volume}{135}}, \bibinfo{pages}{164106}
  (\bibinfo{year}{2011}{\natexlab{a}}).

\bibitem[{\citenamefont{Nicolin and
  Segal}(2011{\natexlab{b}})}]{Nicolin.11.PRB}
\bibinfo{author}{\bibfnamefont{L.}~\bibnamefont{Nicolin}} \bibnamefont{and}
  \bibinfo{author}{\bibfnamefont{D.}~\bibnamefont{Segal}},
  \bibinfo{journal}{Phys. Rev. B} \textbf{\bibinfo{volume}{84}},
  \bibinfo{pages}{161414} (\bibinfo{year}{2011}{\natexlab{b}}).

\bibitem[{\citenamefont{Ruokola and Ojanen}(2011)}]{Ruokola.11.PRB}
\bibinfo{author}{\bibfnamefont{T.}~\bibnamefont{Ruokola}} \bibnamefont{and}
  \bibinfo{author}{\bibfnamefont{T.}~\bibnamefont{Ojanen}},
  \bibinfo{journal}{Phys. Rev. B} \textbf{\bibinfo{volume}{83}},
  \bibinfo{pages}{045417} (\bibinfo{year}{2011}).

\bibitem[{\citenamefont{Segal}(2013)}]{Segal.13.PRB}
\bibinfo{author}{\bibfnamefont{D.}~\bibnamefont{Segal}},
  \bibinfo{journal}{Phys. Rev. B} \textbf{\bibinfo{volume}{87}},
  \bibinfo{pages}{195436} (\bibinfo{year}{2013}).

\bibitem[{\citenamefont{Saito and Kato}(2013)}]{Saito.13.PRL}
\bibinfo{author}{\bibfnamefont{K.}~\bibnamefont{Saito}} \bibnamefont{and}
  \bibinfo{author}{\bibfnamefont{T.}~\bibnamefont{Kato}},
  \bibinfo{journal}{Phys. Rev. Lett.} \textbf{\bibinfo{volume}{111}},
  \bibinfo{pages}{214301} (\bibinfo{year}{2013}).

\bibitem[{\citenamefont{Yang and Wu}(2014)}]{Yang.14.EL}
\bibinfo{author}{\bibfnamefont{Y.}~\bibnamefont{Yang}} \bibnamefont{and}
  \bibinfo{author}{\bibfnamefont{C.}~\bibnamefont{Wu}},
  \bibinfo{journal}{Europhys. Lett.} \textbf{\bibinfo{volume}{107}},
  \bibinfo{pages}{30003} (\bibinfo{year}{2014}).

\bibitem[{\citenamefont{Wang et~al.}(2015{\natexlab{a}})\citenamefont{Wang,
  Ren, and Cao}}]{Wang.15.SR}
\bibinfo{author}{\bibfnamefont{C.}~\bibnamefont{Wang}},
  \bibinfo{author}{\bibfnamefont{J.}~\bibnamefont{Ren}}, \bibnamefont{and}
  \bibinfo{author}{\bibfnamefont{J.}~\bibnamefont{Cao}}, \bibinfo{journal}{Sci.
  Rep.} \textbf{\bibinfo{volume}{5}}, \bibinfo{pages}{11787}
  (\bibinfo{year}{2015}{\natexlab{a}}).

\bibitem[{\citenamefont{Carrega et~al.}(2016)\citenamefont{Carrega, Solinas,
  Sassetti, and Weiss}}]{Carrega.16.PRL}
\bibinfo{author}{\bibfnamefont{M.}~\bibnamefont{Carrega}},
  \bibinfo{author}{\bibfnamefont{P.}~\bibnamefont{Solinas}},
  \bibinfo{author}{\bibfnamefont{M.}~\bibnamefont{Sassetti}}, \bibnamefont{and}
  \bibinfo{author}{\bibfnamefont{U.}~\bibnamefont{Weiss}},
  \bibinfo{journal}{Phys. Rev. Lett.} \textbf{\bibinfo{volume}{116}},
  \bibinfo{pages}{240403} (\bibinfo{year}{2016}).

\bibitem[{\citenamefont{Liu et~al.}(2017)\citenamefont{Liu, Xu, Li, and
  Wu}}]{Liu.17.PRE}
\bibinfo{author}{\bibfnamefont{J.}~\bibnamefont{Liu}},
  \bibinfo{author}{\bibfnamefont{H.}~\bibnamefont{Xu}},
  \bibinfo{author}{\bibfnamefont{B.}~\bibnamefont{Li}}, \bibnamefont{and}
  \bibinfo{author}{\bibfnamefont{C.}~\bibnamefont{Wu}}, \bibinfo{journal}{Phys.
  Rev. E} \textbf{\bibinfo{volume}{96}}, \bibinfo{pages}{012135}
  (\bibinfo{year}{2017}).

\bibitem[{\citenamefont{Wang et~al.}(2017)\citenamefont{Wang, Ren, and
  Cao}}]{Wang.17.PRA}
\bibinfo{author}{\bibfnamefont{C.}~\bibnamefont{Wang}},
  \bibinfo{author}{\bibfnamefont{J.}~\bibnamefont{Ren}}, \bibnamefont{and}
  \bibinfo{author}{\bibfnamefont{J.}~\bibnamefont{Cao}},
  \bibinfo{journal}{Phys. Rev. A} \textbf{\bibinfo{volume}{95}},
  \bibinfo{pages}{023610} (\bibinfo{year}{2017}).

\bibitem[{\citenamefont{May and K\"uhn}(2011)}]{May.11.NULL}
\bibinfo{author}{\bibfnamefont{V.}~\bibnamefont{May}} \bibnamefont{and}
  \bibinfo{author}{\bibfnamefont{O.}~\bibnamefont{K\"uhn}},
  \emph{\bibinfo{title}{Charge and Energy Transfer Dynamics in Molecular
  Systems}} (\bibinfo{publisher}{Wiley-VCH, Weinheim}, \bibinfo{year}{2011}).

\bibitem[{\citenamefont{Majumdar et~al.}(2015)\citenamefont{Majumdar,
  Sierra-Suarez, Schiffres, Ong, Higgs, McGaughey, and Malen}}]{Majumdar.15.NL}
\bibinfo{author}{\bibfnamefont{S.}~\bibnamefont{Majumdar}},
  \bibinfo{author}{\bibfnamefont{J.~A.} \bibnamefont{Sierra-Suarez}},
  \bibinfo{author}{\bibfnamefont{S.~N.} \bibnamefont{Schiffres}},
  \bibinfo{author}{\bibfnamefont{W.-L.} \bibnamefont{Ong}},
  \bibinfo{author}{\bibfnamefont{C.~F.} \bibnamefont{Higgs}},
  \bibinfo{author}{\bibfnamefont{A.~J.~H.} \bibnamefont{McGaughey}},
  \bibnamefont{and} \bibinfo{author}{\bibfnamefont{J.~A.} \bibnamefont{Malen}},
  \bibinfo{journal}{Nano Lett.} \textbf{\bibinfo{volume}{15}},
  \bibinfo{pages}{2985} (\bibinfo{year}{2015}).

\bibitem[{\citenamefont{Tully}(1990)}]{Tully.90.JCP}
\bibinfo{author}{\bibfnamefont{J.~C.} \bibnamefont{Tully}},
  \bibinfo{journal}{J. Chem. Phys.} \textbf{\bibinfo{volume}{93}},
  \bibinfo{pages}{1061} (\bibinfo{year}{1990}).

\bibitem[{\citenamefont{Prezhdo and Kisil}(1997)}]{Prezhdo.97.PRA}
\bibinfo{author}{\bibfnamefont{O.~V.} \bibnamefont{Prezhdo}} \bibnamefont{and}
  \bibinfo{author}{\bibfnamefont{V.~V.} \bibnamefont{Kisil}},
  \bibinfo{journal}{Phys. Rev. A} \textbf{\bibinfo{volume}{56}},
  \bibinfo{pages}{162} (\bibinfo{year}{1997}).

\bibitem[{\citenamefont{Martens and Fang}(1997)}]{Martens.97.JCP}
\bibinfo{author}{\bibfnamefont{C.~C.} \bibnamefont{Martens}} \bibnamefont{and}
  \bibinfo{author}{\bibfnamefont{J.-Y.} \bibnamefont{Fang}},
  \bibinfo{journal}{J. Chem. Phys.} \textbf{\bibinfo{volume}{106}},
  \bibinfo{pages}{4918} (\bibinfo{year}{1997}).

\bibitem[{\citenamefont{Tully}(1998)}]{Tully.98.FD}
\bibinfo{author}{\bibfnamefont{J.~C.} \bibnamefont{Tully}},
  \bibinfo{journal}{Faraday Discuss.} \textbf{\bibinfo{volume}{110}},
  \bibinfo{pages}{407} (\bibinfo{year}{1998}).

\bibitem[{\citenamefont{Kapral and Ciccotti}(1999)}]{Kapral.99.JCP}
\bibinfo{author}{\bibfnamefont{R.}~\bibnamefont{Kapral}} \bibnamefont{and}
  \bibinfo{author}{\bibfnamefont{G.}~\bibnamefont{Ciccotti}},
  \bibinfo{journal}{J. Chem. Phys.} \textbf{\bibinfo{volume}{110}},
  \bibinfo{pages}{8919} (\bibinfo{year}{1999}).

\bibitem[{\citenamefont{Wan and Schofield}(2000)}]{Wan.00.JCP}
\bibinfo{author}{\bibfnamefont{C.}~\bibnamefont{Wan}} \bibnamefont{and}
  \bibinfo{author}{\bibfnamefont{J.}~\bibnamefont{Schofield}},
  \bibinfo{journal}{J. Chem. Phys.} \textbf{\bibinfo{volume}{113}},
  \bibinfo{pages}{7047} (\bibinfo{year}{2000}).

\bibitem[{\citenamefont{Horenko et~al.}(2002)\citenamefont{Horenko, Salzmann,
  Schmidt, and Sch\"utte}}]{Horenko.02.JCP}
\bibinfo{author}{\bibfnamefont{I.}~\bibnamefont{Horenko}},
  \bibinfo{author}{\bibfnamefont{C.}~\bibnamefont{Salzmann}},
  \bibinfo{author}{\bibfnamefont{B.}~\bibnamefont{Schmidt}}, \bibnamefont{and}
  \bibinfo{author}{\bibfnamefont{C.}~\bibnamefont{Sch\"utte}},
  \bibinfo{journal}{J. Chem. Phys.} \textbf{\bibinfo{volume}{117}},
  \bibinfo{pages}{11075} (\bibinfo{year}{2002}).

\bibitem[{\citenamefont{Kelly and Markland}(2013)}]{Kelly.13.JCP}
\bibinfo{author}{\bibfnamefont{A.}~\bibnamefont{Kelly}} \bibnamefont{and}
  \bibinfo{author}{\bibfnamefont{T.~E.} \bibnamefont{Markland}},
  \bibinfo{journal}{J. Chem. Phys.} \textbf{\bibinfo{volume}{139}},
  \bibinfo{pages}{014104} (\bibinfo{year}{2013}).

\bibitem[{\citenamefont{Bai et~al.}(2014)\citenamefont{Bai, Xie, and
  Shi}}]{Bai.14.JPCA}
\bibinfo{author}{\bibfnamefont{S.-M.} \bibnamefont{Bai}},
  \bibinfo{author}{\bibfnamefont{W.-W.} \bibnamefont{Xie}}, \bibnamefont{and}
  \bibinfo{author}{\bibfnamefont{Q.}~\bibnamefont{Shi}}, \bibinfo{journal}{J.
  Phys. Chem. A} \textbf{\bibinfo{volume}{118}}, \bibinfo{pages}{9262}
  (\bibinfo{year}{2014}).

\bibitem[{\citenamefont{Kim and Rhee}(2014)}]{Kim.14.JCP}
\bibinfo{author}{\bibfnamefont{H.~W.} \bibnamefont{Kim}} \bibnamefont{and}
  \bibinfo{author}{\bibfnamefont{Y.~M.} \bibnamefont{Rhee}},
  \bibinfo{journal}{J. Chem. Phys.} \textbf{\bibinfo{volume}{140}},
  \bibinfo{pages}{184106} (\bibinfo{year}{2014}).

\bibitem[{\citenamefont{Wang et~al.}(2015{\natexlab{b}})\citenamefont{Wang,
  Sifain, and Prezhdo}}]{Wang.15.JPCL}
\bibinfo{author}{\bibfnamefont{L.~J.} \bibnamefont{Wang}},
  \bibinfo{author}{\bibfnamefont{A.~E.} \bibnamefont{Sifain}},
  \bibnamefont{and} \bibinfo{author}{\bibfnamefont{O.~V.}
  \bibnamefont{Prezhdo}}, \bibinfo{journal}{J. Phys. Chem. Lett.}
  \textbf{\bibinfo{volume}{6}}, \bibinfo{pages}{3827}
  (\bibinfo{year}{2015}{\natexlab{b}}).

\bibitem[{\citenamefont{Martens}(2016)}]{Martens.16.JPCL}
\bibinfo{author}{\bibfnamefont{C.~C.} \bibnamefont{Martens}},
  \bibinfo{journal}{J. Phys. Chem. Lett.} \textbf{\bibinfo{volume}{7}},
  \bibinfo{pages}{2610} (\bibinfo{year}{2016}).

\bibitem[{\citenamefont{Wang et~al.}(2016)\citenamefont{Wang, Akimov, and
  Prezhdo}}]{Wang.16.JPCL}
\bibinfo{author}{\bibfnamefont{L.~J.} \bibnamefont{Wang}},
  \bibinfo{author}{\bibfnamefont{A.}~\bibnamefont{Akimov}}, \bibnamefont{and}
  \bibinfo{author}{\bibfnamefont{O.~V.} \bibnamefont{Prezhdo}},
  \bibinfo{journal}{J. Phys. Chem. Lett.} \textbf{\bibinfo{volume}{7}},
  \bibinfo{pages}{2100} (\bibinfo{year}{2016}).

\bibitem[{\citenamefont{Agostini et~al.}(2016)\citenamefont{Agostini, Min,
  Abedi, and Gross}}]{Agostini.16.JCTC}
\bibinfo{author}{\bibfnamefont{F.}~\bibnamefont{Agostini}},
  \bibinfo{author}{\bibfnamefont{S.~K.} \bibnamefont{Min}},
  \bibinfo{author}{\bibfnamefont{A.}~\bibnamefont{Abedi}}, \bibnamefont{and}
  \bibinfo{author}{\bibfnamefont{E.~K.~U.} \bibnamefont{Gross}},
  \bibinfo{journal}{J. Chem. Theory Comput.} \textbf{\bibinfo{volume}{12}},
  \bibinfo{pages}{2127} (\bibinfo{year}{2016}).

\bibitem[{\citenamefont{Subotnik et~al.}(2016)\citenamefont{Subotnik, Jain,
  Landry, Petit, Ouyang, and Bellonzi}}]{Subotnik.16.ARPC}
\bibinfo{author}{\bibfnamefont{J.~E.} \bibnamefont{Subotnik}},
  \bibinfo{author}{\bibfnamefont{A.}~\bibnamefont{Jain}},
  \bibinfo{author}{\bibfnamefont{B.}~\bibnamefont{Landry}},
  \bibinfo{author}{\bibfnamefont{A.}~\bibnamefont{Petit}},
  \bibinfo{author}{\bibfnamefont{W.}~\bibnamefont{Ouyang}}, \bibnamefont{and}
  \bibinfo{author}{\bibfnamefont{N.}~\bibnamefont{Bellonzi}},
  \bibinfo{journal}{Annu. Rev. Phys. Chem.} \textbf{\bibinfo{volume}{67}},
  \bibinfo{pages}{387} (\bibinfo{year}{2016}).

\bibitem[{\citenamefont{Aleksandrov}(1981)}]{Aleksandrov.81.ZNA}
\bibinfo{author}{\bibfnamefont{I.~V.} \bibnamefont{Aleksandrov}},
  \bibinfo{journal}{Z. Naturforsch. A} \textbf{\bibinfo{volume}{36}},
  \bibinfo{pages}{902} (\bibinfo{year}{1981}).

\bibitem[{\citenamefont{Gerasimenko}(1982)}]{Gerasimenko.82.TMP}
\bibinfo{author}{\bibfnamefont{V.~I.} \bibnamefont{Gerasimenko}},
  \bibinfo{journal}{Theor. Math. Phys.} \textbf{\bibinfo{volume}{50}},
  \bibinfo{pages}{77} (\bibinfo{year}{1982}).

\bibitem[{\citenamefont{Zhang and Balescu}(1988)}]{Zhang.88.JPP}
\bibinfo{author}{\bibfnamefont{W.~Y.} \bibnamefont{Zhang}} \bibnamefont{and}
  \bibinfo{author}{\bibfnamefont{R.}~\bibnamefont{Balescu}},
  \bibinfo{journal}{J. Plasma Phys.} \textbf{\bibinfo{volume}{40}},
  \bibinfo{pages}{199} (\bibinfo{year}{1988}).

\bibitem[{\citenamefont{Wigner}(1932)}]{Wigner.32.PR}
\bibinfo{author}{\bibfnamefont{E.}~\bibnamefont{Wigner}},
  \bibinfo{journal}{Phys. Rev.} \textbf{\bibinfo{volume}{40}},
  \bibinfo{pages}{749} (\bibinfo{year}{1932}).

\bibitem[{\citenamefont{Kapral}(2015)}]{Kapral.15.JP}
\bibinfo{author}{\bibfnamefont{R.}~\bibnamefont{Kapral}}, \bibinfo{journal}{J.
  Phys.: Condens. Matter} \textbf{\bibinfo{volume}{27}},
  \bibinfo{pages}{073201} (\bibinfo{year}{2015}).

\bibitem[{\citenamefont{Kapral}(2016)}]{Kapral.16.CP}
\bibinfo{author}{\bibfnamefont{R.}~\bibnamefont{Kapral}},
  \bibinfo{journal}{Chem. Phys.} \textbf{\bibinfo{volume}{481}},
  \bibinfo{pages}{77} (\bibinfo{year}{2016}).

\bibitem[{\citenamefont{MacKernan et~al.}(2002)\citenamefont{MacKernan,
  Ciccotti, and Kapral}}]{Kernan.02.JCP}
\bibinfo{author}{\bibfnamefont{D.}~\bibnamefont{MacKernan}},
  \bibinfo{author}{\bibfnamefont{G.}~\bibnamefont{Ciccotti}}, \bibnamefont{and}
  \bibinfo{author}{\bibfnamefont{R.}~\bibnamefont{Kapral}},
  \bibinfo{journal}{J. Chem. Phys.} \textbf{\bibinfo{volume}{116}},
  \bibinfo{pages}{2346} (\bibinfo{year}{2002}).

\bibitem[{\citenamefont{Levitov and Lesovik}(1993)}]{Levitov.93.JETP}
\bibinfo{author}{\bibfnamefont{L.~S.} \bibnamefont{Levitov}} \bibnamefont{and}
  \bibinfo{author}{\bibfnamefont{G.~B.} \bibnamefont{Lesovik}},
  \bibinfo{journal}{JETP Lett.} \textbf{\bibinfo{volume}{58}},
  \bibinfo{pages}{230} (\bibinfo{year}{1993}).

\bibitem[{\citenamefont{Levitov et~al.}(1996)\citenamefont{Levitov, Lee, and
  Lesovik}}]{Levitov.96.JMP}
\bibinfo{author}{\bibfnamefont{L.~S.} \bibnamefont{Levitov}},
  \bibinfo{author}{\bibfnamefont{H.-W.} \bibnamefont{Lee}}, \bibnamefont{and}
  \bibinfo{author}{\bibfnamefont{G.~B.} \bibnamefont{Lesovik}},
  \bibinfo{journal}{J. Math. Phys.} \textbf{\bibinfo{volume}{37}},
  \bibinfo{pages}{4845} (\bibinfo{year}{1996}).

\bibitem[{\citenamefont{Belzig and Nazarov}(2001)}]{Belzig.01.PRL}
\bibinfo{author}{\bibfnamefont{W.}~\bibnamefont{Belzig}} \bibnamefont{and}
  \bibinfo{author}{\bibfnamefont{Y.~V.} \bibnamefont{Nazarov}},
  \bibinfo{journal}{Phys. Rev. Lett.} \textbf{\bibinfo{volume}{87}},
  \bibinfo{pages}{197006} (\bibinfo{year}{2001}).

\bibitem[{\citenamefont{Klich}(2003)}]{Klich.03.NULL}
\bibinfo{author}{\bibfnamefont{I.}~\bibnamefont{Klich}}, in
  \emph{\bibinfo{booktitle}{Quantum Noise in Mesoscopic Physics, NATO Science
  Series II}}, edited by \bibinfo{editor}{\bibfnamefont{Y.~V.}
  \bibnamefont{Nazarov}} (\bibinfo{publisher}{Kluwer, Dordrecht},
  \bibinfo{year}{2003}).

\bibitem[{\citenamefont{Bagrets and Nazarov}(2003)}]{Bagrets.03.PRB}
\bibinfo{author}{\bibfnamefont{D.~A.} \bibnamefont{Bagrets}} \bibnamefont{and}
  \bibinfo{author}{\bibfnamefont{Y.~V.} \bibnamefont{Nazarov}},
  \bibinfo{journal}{Phys. Rev. B} \textbf{\bibinfo{volume}{67}},
  \bibinfo{pages}{085316} (\bibinfo{year}{2003}).

\bibitem[{\citenamefont{Pilgram et~al.}(2003)\citenamefont{Pilgram, Jordan,
  Sukhorukov, and B\"uttiker}}]{Pilgram.03.PRL}
\bibinfo{author}{\bibfnamefont{S.}~\bibnamefont{Pilgram}},
  \bibinfo{author}{\bibfnamefont{A.~N.} \bibnamefont{Jordan}},
  \bibinfo{author}{\bibfnamefont{E.~V.} \bibnamefont{Sukhorukov}},
  \bibnamefont{and}
  \bibinfo{author}{\bibfnamefont{M.}~\bibnamefont{B\"uttiker}},
  \bibinfo{journal}{Phys. Rev. Lett.} \textbf{\bibinfo{volume}{90}},
  \bibinfo{pages}{206801} (\bibinfo{year}{2003}).

\bibitem[{\citenamefont{Saito and Utsumi}(2008)}]{Saito.08.PRB}
\bibinfo{author}{\bibfnamefont{K.}~\bibnamefont{Saito}} \bibnamefont{and}
  \bibinfo{author}{\bibfnamefont{Y.}~\bibnamefont{Utsumi}},
  \bibinfo{journal}{Phys. Rev. B} \textbf{\bibinfo{volume}{78}},
  \bibinfo{pages}{115429} (\bibinfo{year}{2008}).

\bibitem[{\citenamefont{Gutman et~al.}(2010)\citenamefont{Gutman, Gefen, and
  Mirlin}}]{Gutman.10.PRL}
\bibinfo{author}{\bibfnamefont{D.~B.} \bibnamefont{Gutman}},
  \bibinfo{author}{\bibfnamefont{Y.}~\bibnamefont{Gefen}}, \bibnamefont{and}
  \bibinfo{author}{\bibfnamefont{A.~D.} \bibnamefont{Mirlin}},
  \bibinfo{journal}{Phys. Rev. Lett.} \textbf{\bibinfo{volume}{105}},
  \bibinfo{pages}{256802} (\bibinfo{year}{2010}).

\bibitem[{\citenamefont{Esposito et~al.}(2009)\citenamefont{Esposito, Harbola,
  and Mukamel}}]{Esposito.09.RMP}
\bibinfo{author}{\bibfnamefont{M.}~\bibnamefont{Esposito}},
  \bibinfo{author}{\bibfnamefont{U.}~\bibnamefont{Harbola}}, \bibnamefont{and}
  \bibinfo{author}{\bibfnamefont{S.}~\bibnamefont{Mukamel}},
  \bibinfo{journal}{Rev. Mod. Phys.} \textbf{\bibinfo{volume}{81}},
  \bibinfo{pages}{1665} (\bibinfo{year}{2009}).

\bibitem[{\citenamefont{Campisi et~al.}(2011)\citenamefont{Campisi, H\"anggi,
  and Talkner}}]{Campisi.11.RMP}
\bibinfo{author}{\bibfnamefont{M.}~\bibnamefont{Campisi}},
  \bibinfo{author}{\bibfnamefont{P.}~\bibnamefont{H\"anggi}}, \bibnamefont{and}
  \bibinfo{author}{\bibfnamefont{P.}~\bibnamefont{Talkner}},
  \bibinfo{journal}{Rev. Mod. Phys.} \textbf{\bibinfo{volume}{83}},
  \bibinfo{pages}{771} (\bibinfo{year}{2011}).

\bibitem[{\citenamefont{Boudjada and Segal}(2014)}]{Boudjada.14.JPCA}
\bibinfo{author}{\bibfnamefont{N.}~\bibnamefont{Boudjada}} \bibnamefont{and}
  \bibinfo{author}{\bibfnamefont{D.}~\bibnamefont{Segal}}, \bibinfo{journal}{J.
  Phys. Chem. A} \textbf{\bibinfo{volume}{118}}, \bibinfo{pages}{11323}
  (\bibinfo{year}{2014}).

\bibitem[{\citenamefont{Liu and Hanna}(2018)}]{Liu.18.NULL}
\bibinfo{author}{\bibfnamefont{J.}~\bibnamefont{Liu}} \bibnamefont{and}
  \bibinfo{author}{\bibfnamefont{G.}~\bibnamefont{Hanna}}, \bibinfo{journal}{J.
  Phys. Chem. Lett.} \textbf{\bibinfo{volume}{9}}, \bibinfo{pages}{3928}
  (\bibinfo{year}{2018}).

\bibitem[{\citenamefont{Agarwalla et~al.}(2012)\citenamefont{Agarwalla, Li, and
  Wang}}]{Bijay.12.PRE}
\bibinfo{author}{\bibfnamefont{B.~K.} \bibnamefont{Agarwalla}},
  \bibinfo{author}{\bibfnamefont{B.}~\bibnamefont{Li}}, \bibnamefont{and}
  \bibinfo{author}{\bibfnamefont{J.-S.} \bibnamefont{Wang}},
  \bibinfo{journal}{Phys. Rev. E} \textbf{\bibinfo{volume}{85}},
  \bibinfo{pages}{051142} (\bibinfo{year}{2012}).

\bibitem[{\citenamefont{Sergi et~al.}(2003)\citenamefont{Sergi, MacKernan,
  Ciccotti, and Kapral}}]{Sergi.03.TCA}
\bibinfo{author}{\bibfnamefont{A.}~\bibnamefont{Sergi}},
  \bibinfo{author}{\bibfnamefont{D.}~\bibnamefont{MacKernan}},
  \bibinfo{author}{\bibfnamefont{G.}~\bibnamefont{Ciccotti}}, \bibnamefont{and}
  \bibinfo{author}{\bibfnamefont{R.}~\bibnamefont{Kapral}},
  \bibinfo{journal}{Theor. Chem. Acc.} \textbf{\bibinfo{volume}{110}},
  \bibinfo{pages}{49} (\bibinfo{year}{2003}).

\bibitem[{\citenamefont{Sergi and Kapral}(2004)}]{Sergi.04.JCP}
\bibinfo{author}{\bibfnamefont{A.}~\bibnamefont{Sergi}} \bibnamefont{and}
  \bibinfo{author}{\bibfnamefont{R.}~\bibnamefont{Kapral}},
  \bibinfo{journal}{J. Chem. Phys.} \textbf{\bibinfo{volume}{121}},
  \bibinfo{pages}{7565} (\bibinfo{year}{2004}).

\bibitem[{\citenamefont{Kim and Kapral}(2005{\natexlab{a}})}]{Kim.05.JCP}
\bibinfo{author}{\bibfnamefont{H.}~\bibnamefont{Kim}} \bibnamefont{and}
  \bibinfo{author}{\bibfnamefont{R.}~\bibnamefont{Kapral}},
  \bibinfo{journal}{J. Chem. Phys.} \textbf{\bibinfo{volume}{122}},
  \bibinfo{pages}{214105} (\bibinfo{year}{2005}{\natexlab{a}}).

\bibitem[{\citenamefont{Kim and Kapral}(2005{\natexlab{b}})}]{Kim.05.JCPa}
\bibinfo{author}{\bibfnamefont{H.}~\bibnamefont{Kim}} \bibnamefont{and}
  \bibinfo{author}{\bibfnamefont{R.}~\bibnamefont{Kapral}},
  \bibinfo{journal}{J. Chem. Phys.} \textbf{\bibinfo{volume}{123}},
  \bibinfo{pages}{194108} (\bibinfo{year}{2005}{\natexlab{b}}).

\bibitem[{\citenamefont{Esposito and Lindenberg}(2008)}]{Esposito.08.PRE}
\bibinfo{author}{\bibfnamefont{M.}~\bibnamefont{Esposito}} \bibnamefont{and}
  \bibinfo{author}{\bibfnamefont{K.}~\bibnamefont{Lindenberg}},
  \bibinfo{journal}{Phys. Rev. E} \textbf{\bibinfo{volume}{77}},
  \bibinfo{pages}{051119} (\bibinfo{year}{2008}).

\bibitem[{\citenamefont{Thompson and Makri}(1999)}]{Thompson.99.JCP}
\bibinfo{author}{\bibfnamefont{K.}~\bibnamefont{Thompson}} \bibnamefont{and}
  \bibinfo{author}{\bibfnamefont{N.}~\bibnamefont{Makri}}, \bibinfo{journal}{J.
  Chem. Phys.} \textbf{\bibinfo{volume}{110}}, \bibinfo{pages}{1343}
  (\bibinfo{year}{1999}).

\bibitem[{\citenamefont{Wang et~al.}(2001)\citenamefont{Wang, Thoss, and
  Miller}}]{Wang.01.JCP}
\bibinfo{author}{\bibfnamefont{H.}~\bibnamefont{Wang}},
  \bibinfo{author}{\bibfnamefont{M.}~\bibnamefont{Thoss}}, \bibnamefont{and}
  \bibinfo{author}{\bibfnamefont{W.~H.} \bibnamefont{Miller}},
  \bibinfo{journal}{J. Chem. Phys.} \textbf{\bibinfo{volume}{115}},
  \bibinfo{pages}{2979} (\bibinfo{year}{2001}).

\bibitem[{\citenamefont{Dormand and Prince}(1980)}]{Dormand.80.JCAM}
\bibinfo{author}{\bibfnamefont{J.~R.} \bibnamefont{Dormand}} \bibnamefont{and}
  \bibinfo{author}{\bibfnamefont{P.~J.} \bibnamefont{Prince}},
  \bibinfo{journal}{J. Comp. Appl. Math.} \textbf{\bibinfo{volume}{6}},
  \bibinfo{pages}{19} (\bibinfo{year}{1980}).

\bibitem[{\citenamefont{Cuansing and Wang}(2010)}]{Cuansing.10.PRB}
\bibinfo{author}{\bibfnamefont{E.~C.} \bibnamefont{Cuansing}} \bibnamefont{and}
  \bibinfo{author}{\bibfnamefont{J.-S.} \bibnamefont{Wang}},
  \bibinfo{journal}{Phys. Rev. B} \textbf{\bibinfo{volume}{81}},
  \bibinfo{pages}{052302} (\bibinfo{year}{2010}).

\bibitem[{\citenamefont{Liu et~al.}(2018)\citenamefont{Liu, Hsieh, Wu, and
  Cao}}]{Liu.18.JCP}
\bibinfo{author}{\bibfnamefont{J.}~\bibnamefont{Liu}},
  \bibinfo{author}{\bibfnamefont{C.-Y.} \bibnamefont{Hsieh}},
  \bibinfo{author}{\bibfnamefont{C.}~\bibnamefont{Wu}}, \bibnamefont{and}
  \bibinfo{author}{\bibfnamefont{J.}~\bibnamefont{Cao}}, \bibinfo{journal}{J.
  Chem. Phys.} \textbf{\bibinfo{volume}{148}}, \bibinfo{pages}{234104}
  (\bibinfo{year}{2018}).

\bibitem[{\citenamefont{Shi and Geva}(2003)}]{Shi.03.JPCA}
\bibinfo{author}{\bibfnamefont{Q.}~\bibnamefont{Shi}} \bibnamefont{and}
  \bibinfo{author}{\bibfnamefont{E.}~\bibnamefont{Geva}}, \bibinfo{journal}{J.
  Phys. Chem. A} \textbf{\bibinfo{volume}{107}}, \bibinfo{pages}{9059}
  (\bibinfo{year}{2003}).

\bibitem[{\citenamefont{Hanna and Geva}(2008)}]{Hanna.08.JPCB}
\bibinfo{author}{\bibfnamefont{G.}~\bibnamefont{Hanna}} \bibnamefont{and}
  \bibinfo{author}{\bibfnamefont{E.}~\bibnamefont{Geva}}, \bibinfo{journal}{J.
  Phys. Chem. B} \textbf{\bibinfo{volume}{112}}, \bibinfo{pages}{4048}
  (\bibinfo{year}{2008}).

\bibitem[{\citenamefont{Jain and Subotnik}(2018)}]{Jain.18.JPCA}
\bibinfo{author}{\bibfnamefont{A.}~\bibnamefont{Jain}} \bibnamefont{and}
  \bibinfo{author}{\bibfnamefont{J.~E.} \bibnamefont{Subotnik}},
  \bibinfo{journal}{J. Phys. Chem. A} \textbf{\bibinfo{volume}{122}},
  \bibinfo{pages}{16} (\bibinfo{year}{2018}).

\bibitem[{\citenamefont{Nielsen et~al.}(2001)\citenamefont{Nielsen, Kapral, and
  Ciccotti}}]{Nielsen.01.JCP}
\bibinfo{author}{\bibfnamefont{S.}~\bibnamefont{Nielsen}},
  \bibinfo{author}{\bibfnamefont{R.}~\bibnamefont{Kapral}}, \bibnamefont{and}
  \bibinfo{author}{\bibfnamefont{G.}~\bibnamefont{Ciccotti}},
  \bibinfo{journal}{J. Chem. Phys.} \textbf{\bibinfo{volume}{115}},
  \bibinfo{pages}{5805} (\bibinfo{year}{2001}).

\end{thebibliography}

\end{document}